\documentclass[11pt,draftcls,onecolumn]{IEEEtran}

\usepackage{amsmath}
\usepackage{amssymb}
\usepackage{graphicx}
\usepackage{multirow}
\usepackage{amsfonts}
\usepackage{dsfont}

\usepackage{mathrsfs}
\usepackage{color}

\ifCLASSINFOpdf
\else
\fi
\begin{document}
%
% paper title
% can use linebreaks \\ within to get better formatting as desired
% Do not put math or special symbols in the title.
\title{Near-optimal Binary Compressed Sensing Matrix}
%
%
% author names and IEEE memberships
% note positions of commas and nonbreaking spaces ( ~ ) LaTeX will not break
% a structure at a ~ so this keeps an author's name from being broken across
% two lines.
% use \thanks{} to gain access to the first footnote area
% a separate \thanks must be used for each paragraph as LaTeX2e's \thanks
% was not built to handle multiple paragraphs
%

\author{Weizhi~Lu,~Weiyu~Li,~Kidiyo~Kpalma and Joseph~Ronsin
       % John~Doe,~\IEEEmembership{Fellow,~OSA,}
       % and~Jane~Doe,~\IEEEmembership{Life~Fellow,~IEEE}% <-this % stops a space
       \thanks{This paper is developed partially with the results presented at \emph{Data Compression Conference} (\emph{DCC}), Apr.  2012, Salt Lake city, U.S. \cite{Weizhi12}.}
\thanks{ Weizhi Lu, Kidiyo Kpalma and Joseph Ronsin are with Universit\'{e} Europ\'{e}enne de Bretagne (UEB), INSA, IETR,
 UMR CNRS 6164, F-35708 Rennes, France. Email: \{weizhi.lu, kidiyo.kpalma, joseph.ronsin\}@insa-rennes.fr}% <-this % stops a space
\thanks{Weiyu Li is with CREST-ENSAI and IRMAR-Universit\'{e} Rennes 1, 35172 Bruz, France, and  also with  Institute of Mathematics, Shandong University, 25510, Jinan, China. Email:weiyu.li@ensai.fr}% <-this % stops a space
}   %20 avenue des Buttes de Co$\ddot{e}$smes,

\maketitle

% As a general rule, do not put math, special symbols or citations
% in the abstract or keywords.
\begin{abstract}
Compressed sensing is a promising technique  that attempts to faithfully recover sparse signal with as few  linear  and  nonadaptive measurements as possible. Its performance is largely determined by the characteristic of sensing matrix. Recently several   zero-one binary  sensing matrices have been deterministically constructed for their relative low complexity and competitive performance. Considering the   implementation complexity, it is of great practical interest if one could further improve the sparsity of binary matrix without performance loss. Based on the study of restricted isometry property (RIP), this paper proposes the near-optimal  binary  sensing matrix, which guarantees nearly the best performance with as sparse distribution as possible. The  proposed near-optimal binary matrix can be deterministically constructed   with progressive edge-growth (PEG) algorithm. Its performance is confirmed with extensive simulations.

\end{abstract}

% Note that keywords are not normally used for peerreview papers.

% For peer review papers, you can put extra information on the cover
% page as needed:
%\ifCLASSOPTIONpeerreview
 %\begin{center} \bfseries abcdefg \end{center}
%\fi
%
% For peerreview papers, this IEEEtran command inserts a page break and
% creates the second title. It will be ignored for other modes.
\IEEEpeerreviewmaketitle

\section{Introduction}

%\IEEEPARstart{C}ompressed sensing has attracted considerable attention as a potential alternative to Shannon theorem on the sampling of sparse signals. In practice, this technique faces two major challenges. One is the construction of undertermined sensing matrix, which is expected to not only impose weak sparsity constraint on  sensed signals but also hold hardware-friendly structure. The other is the robust recovery of sparse signal only in terms of few linear observations and sensing matrix. For the latter, currently numerous efficient recovery algorithms have been successively proposed along with the study in statistics or approximation theory. However, for the former, although some sensing matrices are also constructed based on some well-known distributions or codes, there is still no explicit way to define or construct the ideal sensing matrix characterized by existing theories (e.g., RIP). In other words, the sensing matrix  optimal both in performance and complexity is still unknown in practice. This paper thus is developed to  address this problem.

\IEEEPARstart{C}ompressed sensing has attracted considerable attention as an alternative to Shannon sampling theorem for the acquisition of sparse signals. This technique includes two challenging tasks. One is the construction of undertermined sensing matrix, which is expected  not only to impose weak sparsity constraint on sensing signals but also to hold low complexity; and the other  is the robust reconstruction of sparse signal over few linear observations. For the latter, currently various optimizing or greedy algorithms of both theoretical and practical performance guarantees have been successively proposed. However, for the former, although a few  sensing matrices have been constructed based on some probability distributions or codes, it is still unknown what kind of matrix is the optimal both in performance and complexity.  This paper  is thus developed to  address this problem.

It is well known that some random matrices generated by certain probabilistic processes, like Gaussian or Bernoulli processes \cite{Candes05} \cite{Candes06near}, guarantee successful signal recovery  with high probability. In terms of complexity,  these dense matrices  are allowed to reduce to more sparse form without obvious performance loss \cite{Achlioptas03} \cite{Baraniuk08} \cite{Berinde08}.  However, they are still impractical due to randomness. In this sense, it is of practical importance to explore  deterministic sensing matrices of both favorable performance and feasible structure. Recently several deterministic  sensing matrices have been sequentially proposed base on some families of codes, such as BCH codes \cite{Ailon08} \cite{Amini11}, Reed-Solomon codes \cite{Akcakaya08}, Reed-Muller codes \cite{Howard08} \cite{Calderbank10B} \cite{Calderbank10}, LDPC codes \cite{Pham09} \cite{Barg10}, etc. These codes are exploited based the fact that coding theory attempts to maximize the distance between two distinct codes, while in some sense this rule  is also preferred for compressed sensing that tends to minimize the correlation between distinct columns of a sensing matrix \cite{Pham09} \cite{Amini11}.   From the viewpoint of application, it is interesting to know which kind of deterministic matrix is the best in performance. Unfortunately, to the best of our knowledge, there is still no impressively theoretical  works covering this problem.

%DeVore \cite{DeVore07} theoretically defines the best known performance for deterministic matrix with the polynomial over Finite Field. However, the guaranteed best performance is still worse than Random matrices, and further, the explicit construction of large size matrix is also unknown.

%(In spite of explicit structure, these matrices still suffer from uncertain RIP and unstable performance \cite{Lidandan12}.)  For instance, LDPC code, vandermonde code,Obviously they are welcomed for hardware-implementation for their $\{0,1\}$ form as well as comparable performance with Gaussian matrix.

Note that the aforementioned deterministic sensing matrices mainly take entries from    bipolar set $\{-1,1\}$, ternary set $\{0,\pm1\}$, or binary set $\{0,1\}$. As there is no  matrix reported to obviously outperform others, it is practically preferable to exploit the one with   lowest complexity. Therefore, in this paper we are concerned only with  hardware-friendly $\{0, 1\}$ binary matrix, and aim at maximizing its sparsity at least without performance loss.   Note that the deterministic binary matrices based on codes \cite{Amini11}  \cite{Indyk08} are not very sparse in structure, since statistically they should take 0 and 1 with equal probability. However, the well-known deterministic binary matrix generated with the polynomials in finite fields of size $p$, is relatively sparse with the proportion of nonzero entries being $1/p$ in each column \cite{DeVore07}. But up to now there is still few knowledge about the  practical performance of this kind of matrix, i.e., the performance over both the varying order of polynomials and the varying size of finite field. The studies on expander graph \cite{Xu07} \cite{Jafarpour09} also proposed deterministic performance guarantees for sparse binary matrix, while the practical construction of the desired  matrix is still a challenging task. Recently the sparse binary parity-check matrix of LDPC codes drew our attention for its high sparsity and favorable performance \cite{Dimarkis12} \cite{Lidandan12} \cite{LiuXinji13}. This type of matrices enjoys much higher sparsity than others, e.g., empirically only about $3$ nonzero entries are required for each column  of 'good' LDPC codes. Nevertheless,  their performance cannot be ensured to be the best for compressed sensing. Clearly it is hard to determine the optimal binary matrix both in performance and sparsity only with aforementioned works. An interesting question then arises:  does there exist some optimal distribution for binary sensing matrix such that it could achieve the best performance with as high sparsity as possible? Inspired by the  graph-based analysis method for sparse binary matrix \cite{Tanner81} \cite{Tanner01} \cite{Sipser96}, this paper successfully determines the near-optimal distribution of binary sensing matrix. The proposed approach proceeds into two steps: first, the binary matrix is   categorized  into two types in terms of  graph structure, and then the sparsity of near-optimal binary sensing matrix is derived by evaluating the restricted isometry property (RIP).

The rest of the paper is organized as follows.  In the next Section, we provide the fundamental knowledge about compressed sensing as well as the  binary matrix characterized with bipartite graph. In section \uppercase\expandafter{\romannumeral3}, the binary matrix is divided into two types in terms of  graph structure, and then the near-optimal sensing matrix is derived by analyzing their  RIP.  In Section \uppercase\expandafter{\romannumeral4}, the proposed near-optimal matrix is deterministically constructed with progressive edge-growth (PEG) algorithm, and its performance is  confirmed by performing extensive comparisons with other matrices. Finally, this paper is concluded in Section \uppercase\expandafter{\romannumeral5}. To make the paper more readable, several long proofs are presented in  a series of appendices.

\section{Preliminaries}

\subsection{Compressed sensing}

Suppose that a $k$-sparse signal $x\in \mathbb{R}^N$ with at most \emph{k} nonzero entries, is sampled by an undetermined matrix $A \in \mathbb{R}^{M\times N}$ with $M<< N$ as follows
\begin {equation}
y=Ax.
\end {equation}
Compressed sensing  asserts that  $x$ could be perfectly recovered from a low-dimensional observation $y \in \mathbb{R}^M$,  if the sensing matrix $A$ satisfies  RIP \cite{Candes05}. The solution to formula (1) is  customarily formulated as an $\ell_1$-regularized minimization  problem

 \begin{equation}
 \min ||\hat{x}||_1 ~~\text{subject~to}~~ y=A\hat{x},
 \end{equation}
which could be well solved or approximated by numerous algorithms as reviewed in \cite{Sturm11}.

 Prior to introducing RIP, we have to review  a term called $k$-restricted isometry constant (RIC), denoted as $\delta_k$, which is the smallest quantity obeying
\begin{equation}
(1-\delta_k)||x_T||^2\leq  ||A_T x_T||^2 \leq (1+\delta_k)||x_T||^2
\end{equation}
for arbitrary submatrix $A_T\in \mathbb{R}^{M\times|T|}$ and corresponding vector $x_T\in \mathbb{R}^{|T|}$ under $|T|\leq k$, where $T\subset \{1,...,N\}$ denotes the column index subset of $A$, and $|T|$ is its  cardinality. Then RIP is stated by asserting that a $k$-sparse signal can be recovered faithfully with formula (1), if $\delta_{2k}$ is less than some given threshold.  In practice, to recover $x$ with large $k$, compressed sensing obviously requires that $\delta_{2k}$ is as small as possible. Equivalently,  Gramian matrix $A_T'A_T$ is preferred to approximate isometry as $|T|$ increases, where $A_T'$ is the transpose of $A_T$. %Consequently, roughly speaking, compressed sensing motivates us to construct the measurement matrix with the goal of maximizing    $k$ and minimizing $\delta_{k}$.

\subsubsection{Solution to RIP}
In practice, to evaluate a given sensing matrix, it is crucial to derive the largest $k$ with $\delta_{2k}$ bounded by RIP. Theoretically, the solution to  the RIC of $A_T'A_T$, can be transformed to the pursuit for the extreme eigenvalues of $A_T'A_T$, since
\begin{equation}
1-\delta_k \leq \lambda_{k}\leq \frac{x_T'A_T'A_Tx_T}{x_T'x_T}\leq\lambda_{1} \leq 1+\delta_k
\end{equation}
where $\lambda_1$ and  $\lambda_k$ represent the two extreme eigenvalues of $A_T'A_T $. For notational convenience, in the following part we  let $|T|=k$, though $|T|\leq k$ is used in the former definition of RIC.   Clearly, for a given sensing matrix, the solution to  the extreme eigenvalues of  $A_T'A_T $  is NP-hard \cite{Needell10} \cite{Tillmann12} \cite{Bandeira12}. In practice, this problem tends to be tackled by analyzing the distribution of the elements of $A_T'A_T $, by regrading it as a \emph{random } symmetric matrix since practically the combinatorial number of  the subset $T$  is likely to be very large  \cite{Blanchard11}. As it is known, Wigner semicircle law \cite{Pastur72} is suitable for bounding the extreme eigenvalues of random symmetric matrix \cite{Gurevich08}. However, this algorithm presents an obvious drawback,  its solution accuracy could  be ensured only when the size of  $A_T'A_T $  gets close to infinity. This is contradictory to the fact that RIP is preferred to be accurately derived as $|T|$ is relatively small, especially when the size of sensing matrix is not large enough. Gershgorin circle theorem \cite{Horn85} is also a popular solution algorithm for the  eigenvalues of square matrix. Similar with Wigner semicircle law, this algorithm also suffers from  inaccuracy. Exactly speaking, with Gershgorin circle theorem, it can be observed that the bound of any eigenvalue of binary matrix can  be achieved only when  the following two conditions are simultaneously satisfied: 1) the nonzero entries of the eigenvector share the same magnitude; 2) the elementwise products between the eigenvector and the off-diagonal elements of the corresponding matrix row vector  should hold the same sign. Obviously it seems hard to ensure that actual matrices fulfill these two conditions. Furthermore, for a square matrix of given distribution, it is unknown to what extent  two previous conditions can be satisfied, such that one cannot intuitively judge the accuracy of the bounds derived with  Gershgorin circle theorem.

%In this sense, this paper exploits a more practical algebra algorithm \cite{zhan06} rather than semicircle law to derive the RIP, which theoretically can achieve the extreme eigenvalues of random symmetric matrix of arbitrary size, if the element distribution of random matrix is ideal. For details, the readers are referred to  the following proofs for RIPs where some specific distributions required for achieving the bounds of eigenvalues are illustrated.  Here, it should be noted that in the following study, the matrices we  define cannot be ensured to definitely hold the distributions that achieve the bounds of extreme eigenvalues, due to the insufficient description for the  distribution of $A_T'A_T $. But considering this paper aims to compare RIPs rather than derive accurate RIP, this inherent deficiency is hoped to impose little negative influence on our results.

Based on the above observations, this paper exploits a more practical algebra algorithm \cite{zhan06} to explore the extreme eigenvalues of  $A_T'A_T $. This algorithm  can accurately bound the extreme eigenvalues of random symmetric matrix of arbitrary size, under the assumption that random matrix $A_T'A_T $ could achieve some specific distribution which will be detailed in the next Section.  Of course, such algorithm is also imperfect, because the required specific distributions for the extreme eigenvalues seem hard to be satisfied for all actual sensing matrices. However, for a given sensing matrix, the accuracy of the solution allows to be intuitively judged, since the accuracy depends on the  the distribution of $A_T'A_T $ while in practice the distribution usually could   be  characterized. This is also one advantage of the adopted algorithm  \cite{zhan06} over Wigner semicircle law and Gershgorin circle theorem.

 %It is necessary to note that we cannot ensure that the practical matrix definitely could   achieve the ideal distribution required for the bounds of the extreme eigenvalues derived by the algorithm in \cite{zhan06}, and this reveals that the theoretical bounds of RIP we will derive are possibly  inaccurate for some real matrices. In addition, as it is known, it is not very rigorous to judge  the performance of given sensing matrix with the RIP, which in fact  is only a sufficient condition for compressed sensing. Despite two inevitable theory deficiencies above, the near-optimal binary matrix we will define indeed shows expected performance in the final simulation.

%Considering two inevitable theory deficiencies above, the near-optimal matrix we will theoretically propose is further confirmed with extensive simulations.

 \subsection{Binary matrix characterized with bipartite graph}

\begin{figure}[t]

\centering
\graphicspath{fig}
\includegraphics[width=0.8\textwidth]{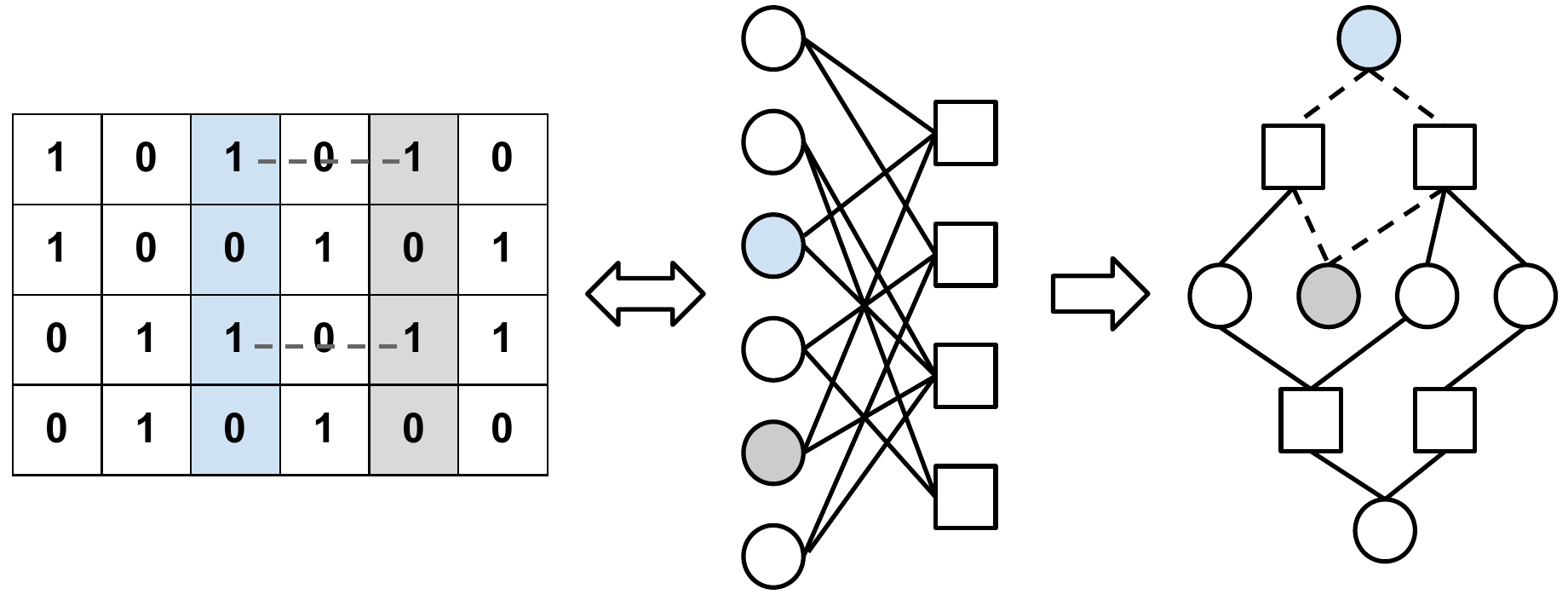}

\caption{From left to right: binary matrix, bipartite graph and  subgraph. Variable node and measurement node are denoted with circle and square, respectively. If two variable nodes share two same nonzero positions in their corresponding columns of binary matrix, they will form a shortest cycle  of length 4 (dashed lines) in the  subgraph expanded from each of them.}

\end{figure}

 In this paper, we mainly study the \emph{regular}  binary matrix, which has the same number of nonzero entries  in columns/rows. For notational simplicity, in the following work  regular binary  matrix is  called  binary matrix except for specific explanation. A given binary matrix can be uniquely associated with a bipartite graph, which consists of two classes of nodes, customarily called variable nodes and  measurement nodes, corresponding to the columns and  rows, respectively. A pair of variable and measurement nodes is connected by an edge if  binary matrix has nonzero entry in the corresponding position.   Hence, one can expand a subgraph from each variable node to connected nodes through edges. A closed path in subgraph is called a \emph{cycle}. The length of the shortest cycle among all subgraphs is defined as  the \emph{girth} of the bipartite graph. One can derive that the value of the girth is even and not less than 4. For better understanding, we give an example in Figure 1. Note that sensing matrix is typically required to be normalized with columns. In this paper, assume that binary matrix has degree $d$, namely holding $d$ nonzero entries in each column, the nonzero entries are thus set to  $1/\sqrt{d}$ instead of $1$.

  To explore the potential sparsest sensing matrix, here we propose two critical definitions as shown in Definitions $1$ and $2$, which  categorize the binary matrices  into two classes in terms of girth distribution. Note that, the binary matrix with $g> 4$ is preferred for LDPC codes as parity-check matrix,  so its construction has been extensively studied in practice. In contrast, for the binary matrix with $g=4$, there is still no explicit way to construct a binary matrix with  a given maximum correlation $s/d$, $2\leq s\leq d-1$. But recall that the nomarlized random binary matrix with uniform degree $d$, denoted as $R(M,N,d)$,  is a typical  binary matrix with $g=4$ while without specific constraint on the maximum correlation $s/d$. So in the following study, it is exploited as a practical version of the binary matrix with $g=4$.

\emph{\textbf{Definition 1} (Binary matrix with girth $g>4$)}:~~A binary matrix, denoted as $A(M,N,d) \in \{0,1/\sqrt{d}\}^{M\times N}$, consists of $2\leq d\leq M-2$ nonzero entries per column and $Nd/M$ nonzero entries per row. In the associated bipartite graph, the girth is required to be larger than $4$. Equivalently,  any two distinct columns of this matrix are allowed to share at most one same nonzero position.
\vspace{3pt}

\emph{\textbf{Definition 2} (Binary matrix with girth $g=4$)}:~~A binary matrix, denoted as $A(M,N,d,s) \in \{0,1/\sqrt{d}\}^{M\times N}$, consists of $3\leq d\leq M-2$ nonzero entries per column and $Nd/M$ nonzero entries per row.  In the associated bipartite graph, the girth takes value 4. Accordingly, the maximum correlation value between two distinct columns  is $s/d$ with $2\leq s\leq d-1$.

It is known that the maximum correlation between distinct columns, denoted as $\mu$, has been a basic performance indicator for the sensing matrix \cite{Donoho11}. So in the following Lemmas 1 and 2, we derive the correlation distributions of the binary matrices with $g>4$ and with $g=4$ (random binary matrix), respectively. It can be observed that the correlation distribution of the binary matrix with $g>4$ is simply binary, while the distribution of random binary matrix is relatively complicated. With the  formula (7)  for random binary matrix, it can be deduced that the probability of taking correlation value $s/d$ will significantly decrease as $s$ increases under the condition of $M\gg d$. This reveals that $\mu$  probably takes values much less than $1$, i.e. $s<<d$, such that  the practical random binary matrix with limited columns has no same columns, . With the correlation characters shown in Lemmas 1 and 2, and the law that smaller $\mu$ leads to larger $k$ \cite{Donoho11}:
 \begin{equation}
 k<\frac{1}{2}(1+1/\mu),
 \end{equation}
  it is reasonable to expect that the binary matrix with $g>4$  probably approaches the best sensing performance, as $d$ achieves its upper bound.  In the next Section, we further confirm this conjecture with RIP analysis.
\vspace{3pt}

\textbf{Lemma 1 (Correlation distribution of binary matrix with $g> 4$)}:~~ Any two distinct columns of binary matrix $A(M,N,d)$ with $g>4$  take correlation values as

\begin{equation}
a_i'a_{j,j\neq i}=
  \left\{
   \begin{array}{cl}
   \textstyle 1/d&\textstyle \text{with probability} ~\rho=\frac{Nd^2-Md}{(N-1)M}  \\
 \textstyle 0& \textstyle \text{with probability} ~1- \rho \\

   \end{array}
  \right.
  \end{equation}
where $a_i$ and $a_j$ denote two distinct columns of $A(M,N,d)$.
\vspace{3pt}
%\noindent\emph{Proof.}
\begin{IEEEproof}In bipartite graph associated with $A(M,N,d)$, any variable node $v_i$, $i\in \{1,...,N\}$, holds $d$ neighboring measurement nodes $c_{b_k}$, where the subscript $b_k\in C\subset\{1,...,M\}$ denotes the index of measurement node,  $k\in \{1,...,d\}$, $|C|=d $;  each measurement node $c_{b_k}$ further connects with other $\frac{Nd}{M}-1$ variable nodes $v_j$, where $j\in V_{b_k}\subset \{1,...,N\} \setminus i$ represents the index of variable node, $|V_{b_k}|=\frac{Nd}{M}-1$. Since variable node $v_i$ has girth $g>4$, we have $V_{b_e}\bigcap V_{b_f}=\emptyset$, and then derive $|V_{b_1}\bigcup V_{b_2}\bigcup ...\bigcup V_{b_k}|=d\times (\frac{Nd}{M}-1)$, where $e,f \in \{1,...,k\}$ and $e\neq f$. Therefore, among $N-1$ variable nodes, there are $\frac{Nd^2-Md}{M}$  connected to variable node $v_i$ through one measurement node. This reveals that any column of $A(M,N,d)$  has $\frac{Nd^2-Md}{M}$ correlated columns with correlation value $1/d$. Then the probability that any two distinct columns correlate to each other is derived as $\frac{Nd^2-Md}{(N-1)M}$ .
 \end{IEEEproof}

\vspace{5pt}
\textbf{Lemma 2 (Correlation distribution of random binary matrix with $g=4$)}:~~ Any two distinct columns of random binary matrix $R(M,N,d)$  take correlation values as
\begin{equation}
\displaystyle  r_i'r_{j,j\neq i}= s/d~~~\text{with probability} ~\textstyle \rho=\frac{d!d!(M-d)!(M-d)!}{(d-s)!(d-s)!s!(M-2d+s)!M!}
  \end{equation}
where $r_i$ and $r_j$ denote two distinct columns, and $0\leq s \leq d$.
\vspace{3pt}
%\noindent\emph{Proof.}
\begin{IEEEproof} The correlation between columns is determined by the overlap rate of  nonzero positions of two columns. Assume that two columns have $s$ same nonzero positions, $0\leq s \leq d$, then the corresponding probability can be easily derived as
$$
\textstyle
\rho=\frac{C_M^{d-s}C_{M-(d-s)}^{d-s}C_{M-2(d-s)}^s}{C_M^dC_M^d}=\frac{d!d!(M-d)!(M-d)!}{(d-s)!(d-s)!s!(M-2d+s)!M!}
$$
if $d$ nonzero positions are selected randomly and uniformly in each column of $R(M,N,d)$.
 \end{IEEEproof}

%This means the ideal LDPC matrix with $g=4$ described in the following paper possibly cannot be practically constructed.
%Except for special explanation, in the following part 'LDPC matrix' simply represents the LDPC matrix with $g>4$ .

%This section defines one class of regular LDPC matrix $A(M,N,d)$ with \emph{girth} larger than 4 in \emph{Definition 1}. Subsequently, two related concepts: \emph{Tanner graph} and \emph{girth} (abbreviated as $g$),  are introduced in \emph{Definitions} \emph{2} and \emph{3}. Note that, to express the variation of correlation between columns as $d$ varies, in this paper the columns of LDPC matrix are normalized and  nonzero entries are assigned to $1/\sqrt{d}$ instead of $1$. Theoretically,The normalization  will not change the RIP.  Except for special explanation, in the following part LDPC matrix only represents the matrix in \emph{Definition 1}. %In theory, the RIP of binary LDPC matrix is independent of the value of nonzero entry.  This property is helpful to estimate the expectation of RIP, which is more useful in application.  , which may be better understood in Figure 1

\section{Near-optimal binary matrix for compressed sensing}

In this section, the RIPs of binary matrices with $g>4$ and $g=4$ are first evaluated in Theorems $1$-$3$, and then the near-optimal binary matrix is derived with Theorem $4$ and related remarks.

%In this sense, the accuracy of the solution algorithm to RIP
\subsection{RIP of binary matrix with girth larger than 4}

%\begin{figure}[t]

%\centering
%\graphicspath{fig}
%\includegraphics[width=0.5\textwidth]{fig/fdelta_k}

%\caption{The statistical result for RIC-$\delta_k$ of LDPC matrix $A(200,400,7)$ through $10^3$ simulations, compared to the theoretical results from RIP-1 and RIP-2.}

%\end{figure}
As sated before,  RIP can be derived by searching the extreme eigenvalues of random symmetric matrix $A_T'A_T$ with arbitrary $T \subset\{1,...,N\}$. In terms of Lemma 1 and the normalization of columns,  we can easily derive that  $A_T'A_T\in \{0,1,1/d\}^{k\times k}$ has the diagonal equal to 1, and the corresponding off-diagonal holds binary  distribution as shown in Lemma 1. With above given distribution, the extreme eigenvalues of $A_T'A_T$  can be derived according to the algebraic algorithm \cite{zhan06}. Then the RIP is derived from Theorem 1.

\vspace{3pt}
\textbf{Theorem 1 (RIP-1)}:~~ The binary matrix $A(M,N,d)$ with $g>4$ satisfies RIP with
\begin{equation}
\delta_k = \frac{3k-2}{4d+k-2}.
\end{equation}
%It follows from $\delta_{2k}<0.4652$  \cite{Foucart10} that  $k<0.3671d+0.2110$ ensures the faithful recovery based on $\ell_1$-minimization.
%\noindent\emph{Proof.} Please see Appendix \uppercase\expandafter{\romannumeral1}
\vspace{2pt}
\begin{IEEEproof} Please see Appendix A.\end{IEEEproof}
\vspace{3pt}

%\noindent{\emph{Remarks:}}

\textbf{Remark:} From the proof of Theorem 1, it can be observed that the  bounds of the two extreme eigenvalues  are  achieved only on the condition that the proportion \emph{p} of nonzero entries in the off-diagonal of $A_T'A_T$,  could take value 1 or 0.5, for any $|T|$.
  However, as Lemma  1  discloses, this condition cannot be satisfied all the time, because  with high probability  the proportion  $p$ should center on $\rho<1$  as $|T|$ increases. This is demonstrated by a real example in Figure 2, which shows the simulation results from a  binary matrix $A(200,400,7)$ with $g>4$  constructed with PEG algorithm. As can be seen in Figure 2, the corresponding proportion $p$  will rapidly converge to the theoretical value $\rho=0.2281< 0.5$,  as $|T|$ increases. Therefore, for a large size binary matrix with  $k$ large enough such that $p=\rho$, it is preferable to derive RIP with Wigner semicircle law \cite{Pastur72}, if the condition of $|T|\rightarrow \infty$ could also be approximately satisfied. The  related RIP-2 is  provided in  Theorem 2. Note that, to obtain a relatively fair comparison, in the next Section we only adopt  RIP-1 and RIP-3 which are derived with the same solution algorithm \cite{zhan06}.

   %Note that,  Theorem 2 is only feasible for binary matrix with  RIP of order $k$ large enough such that the condition for semicircle law, $|T|\rightarrow \infty$, could be well satisfied.   Considering generality and accuracy, RIP-1 rather than RIP-2 is used in the following comparison between RIPs.  %In contrast, RIP is more interested with small $|T|$.
    %In this sense, we prefer to  use RIP-1 for the following comparison between RIPs.

\begin{figure}[t]

\centering
\graphicspath{fig}
\includegraphics[width=0.495\textwidth]{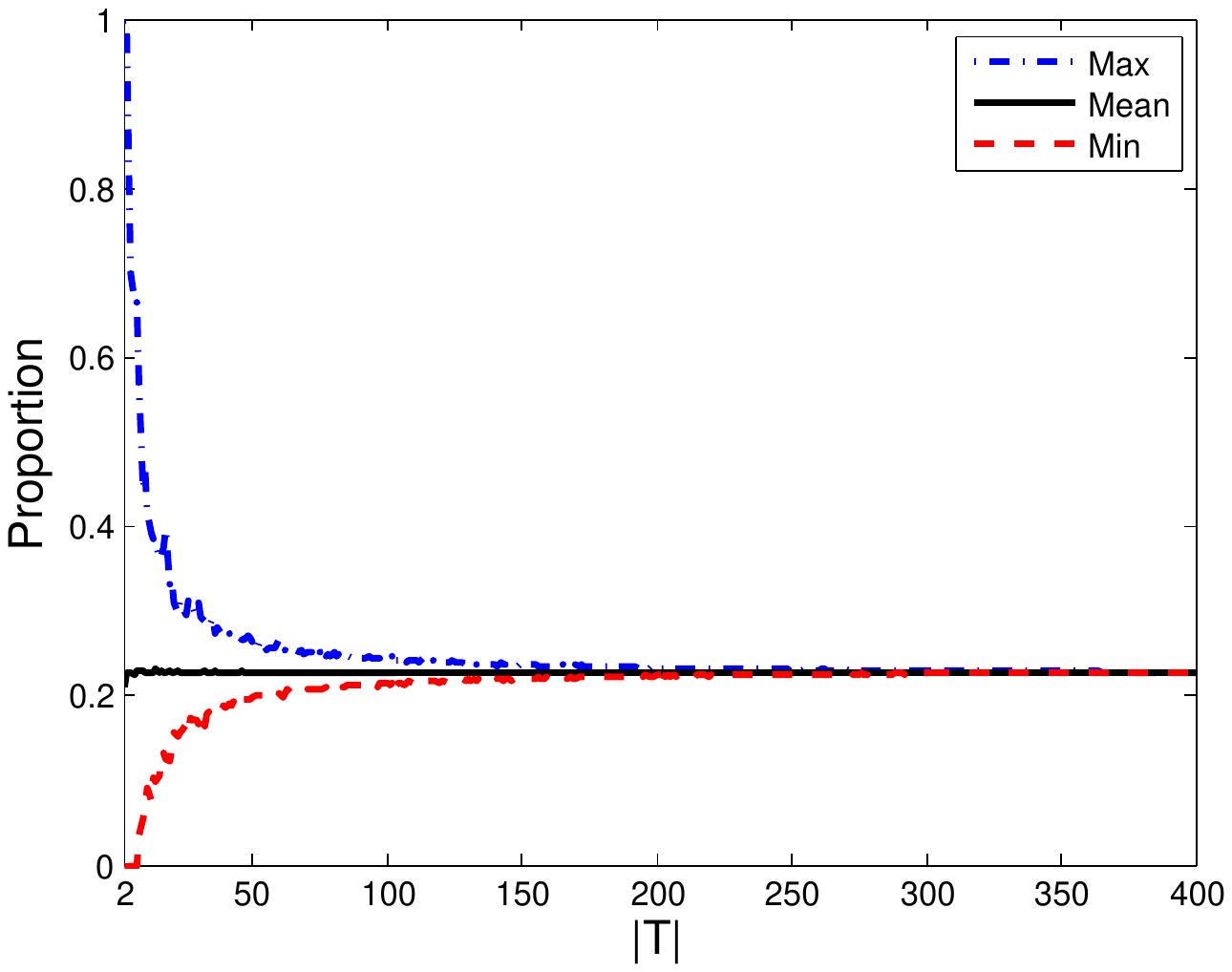}
\includegraphics[width=0.495\textwidth]{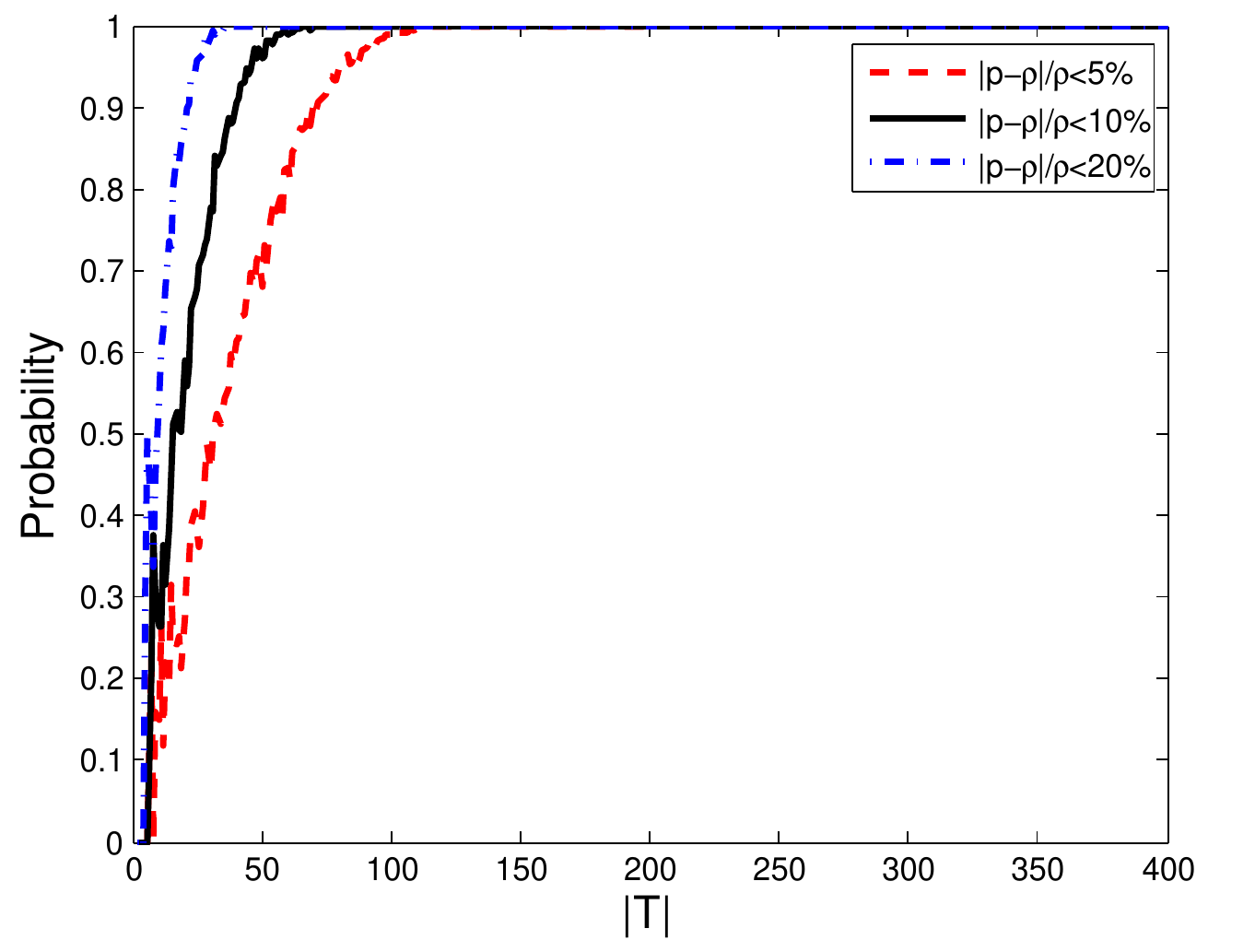}\\
(a)~~~~~~~~~~~~~~~~~~~~~~~~~~~~~~~~~~~~~~~~~~~~~~~~~~~~~~~~~~~~~~(b)\\
\caption{(a) Maximum, mean and minimum of the proportion $p$ of nonzero entries in the off-diagonal of $A'_TA_T$, for a real binary matrix $A(400,200,7)$ of $g>4$ constructed with PEG. As expected, the mean is equal to the theoretical value $\rho=0.2281$ derived with  Lemma 1. (b) The probability that the proportion $p$  centers on the theoretical value $\rho$ within error bound $|p-\rho|/\rho$. Each point of (a) and (b) is derived  from $10^3$ simulations.}

\end{figure}

 % m Since we cannot develop a more accurate mathematic model for RIP by involving the distribution of nonzero entries disclosed in \emph{lemma 1}.
\vspace{3pt}
\textbf{Theorem 2 (RIP-2)}:~~ Assume that the off-diagonal elements of $A_T'A_T$ take nonzero values with probability $\rho=\frac{Nd^2-Md}{(N-1)M}$ while $|T|(|T|-1)\rho\geq 2$, then the RIC  of $A(M,N,d)$ can be approximately formulated as
\begin{equation}
\delta_k=\frac{k\rho+2\sqrt {k\rho(1-\rho)}+1}{k\rho-2\sqrt {k\rho(1-\rho)}+2d+1}
\end{equation}  if $k=|T|\rightarrow\infty$. %Then roughly derive $k<$ for faithful recovery based on $l_1$ constraints .

\begin{IEEEproof}  Please see Appendix B. \end{IEEEproof}

\subsection{RIP of binary matrix with girth equal to 4}
With the definition of binary matrix $A(M,N,d,s)$ with $g=4$, it is easy to derive that $A_T'A_T$ has the maximum correlation of $\mu=s/d$, and the off-diagonal elements possibly drawn  from the set $\{0,1/d,...,s/d\}$, where  $3 \leq d\leq M-2$ and $2\leq s \leq d-1$.  By proceeding the proof similar with  Theorem 1, the RIP of binary matrix with $g=4$ is obtained in Theorem 3.

%\noindent\emph{Proof.} Please see Appendix \uppercase\expandafter{\romannumeral2}

\vspace{3pt}
\textbf{Theorem 3 (RIP-3)}:~~The binary matrix $A(M,N,d,s)$ with $g=4$ and  $\mu=s/d$ , where $2\leq s\leq d-1$ and $3\leq d\leq M-2$, satisfies RIP with

 \begin{equation}
 \textstyle \delta_k =
 \left\{
   \begin{array}{ll}
\Large{\frac{\textstyle (3k-2)s}{\displaystyle (k-2)s+4d}} & \footnotesize {if~~ 3\leq d\leq \frac{M}{2}~ ~and~~ 2\leq s\leq d-1} \\[10pt]
\Large{\frac{\textstyle (3k-2)s+(k-2)(M-2d)}{\displaystyle (k-2)s-(M-2d)k+2M}} & \footnotesize{if~ ~\frac{M}{2}< d\leq M-2~ ~and~ ~2d-M\leq s\leq d-1}\\
   \end{array}
   \right.
 \end{equation}

\vspace{10pt}
\begin{IEEEproof}  Please see Appendix C.\end{IEEEproof}

\vspace{3pt}

%\noindent{\emph{Remarks:}}
\textbf{Remark:} Similar with RIP-1, the bounds of the two extreme eigenvalues for RIP-3 are  achieved as the off-diagonal elements of $A_T'A_T$ can  take the maximum nonzero value $s/d$  with probability 1, or take binary value $\{0, s/d\}$ with equal probability. Unfortunately,  two previous conditions cannot be ensured by a practical matrix which possibly takes $\mu=s/d$ with a relative low probability as Lemma 2 shows. This means that the RIP-3 cannot reasonably describe the RIP of the binary matrix that  takes nonzero correlation values $1/d$  with high probability rather than $\{2/d,...,s/d\}$. In this case, its RIP is more close to RIP-1.  Due to the inaccuracy of RIP-3,  we have to specifically discuss this case in the following pursuit of the best sensing matrix.

\subsection{Near-optimal binary sensing matrix}

%According to the definition of binary matrix $A(N,M,d)$ with $g>4$, there is an upper bound, denoted as $d_{max}$,  for the degree $d$,  which can be approximated as  $d_{max}<\sqrt{M}<M/2$  with the inequality   $1+d(\frac{dN}{M}-1)\leq N$ shown in  \cite{Huxiaoyu05}. This subsection discloses that the binary matrix $A(N,M,d_{max})$ probably possesses nearly the best RIP performance with the following Theorem 4 as well as related remarks.

  According to the definition of binary matrix $A(N,M,d)$ with $g>4$, there is  for the degree $d$, an upper bound denoted as $d_{max}$,  which can be approximated as  $d_{max}<\sqrt{M}<M/2$  with the inequality   $1+d(\frac{dN}{M}-1)\leq N$ shown in  \cite{Huxiaoyu05}. From the Theorem 4 and related remarks, it can be observed that in theory there possibly exist three classes of binary matrices holding better RIP than the matrix $A(N,M,d_{max})$, while in fact only one of them can be practically constructed with slightly better RIP. Therefore, in this paper the matrix $A(N,M,d_{max})$ is viewed as the 'near-optimal' binary sensing matrix, because it achieves nearly the best RIP with as sparse distribution as possible.

\textbf{Theorem 4 (RIP of binary matrix $A(N,M,d_{max})$ )}:~~ Among all binary matrices with  $g\geq 4$ and $d\leq d_{max}$, the binary matrix $A(N,M,d_{max})$ holds the best  RIP.
Compared to the  binary matrices with $g=4$ and $d> d_{max}$, the binary matrix $A(N,M,d_{max})$ also performs better  under each of the following two  conditions:
\begin{itemize}
\item [1)] $d_{max}\geq d/s$, ~if $d_{max} <d \leq M/2$
\item [2)] $d_{max} \geq \frac{(k+1)(2d-M)}{6s+2(2d-M)}$,~  if $M/2 <d \leq M-2$
\end{itemize}
where $d$, $s$ and $k$ follow the definitions of Theorems 1 and 3.

%There exists an upper bound, denoted as $d_{max}$, for the degree $d$ of binary matrix $A(N,M,d)$  with $g>4$, which can be approximated as  $d_{max}<\sqrt{M}<M/2$  in terms of $1+d(\frac{dN}{M}-1)\leq N$  \cite{Huxiaoyu05}. It is asserted that $A(N,M,d_{max})$ possesses nearly the best RIP performance among all binary matrices. This reveals that there exists a nearly optimal degree/sparsity distribution for binary sensing matrix.

\vspace{4pt}
\begin{IEEEproof} We first  prove that $A(M,N,d_{max})$ is the best one among all binary matrices with $g>4$, namely $A(M,N,d_{max})$ is better than $A(M,N,d)$ with $d\leq d_{max}$.
With RIP-1, it is easy to derive that the RIC-$\delta_k$  decreases as  the degree $d$ increases. Thus, it is proved that  binary matrix with  $g>4$  achieves best RIP as  $d=d_{max}$.

 Then, we are ready to prove that $A(M,N,d_{max})$ is still better than the binary matrix with $g=4$. Note that,   the binary matrix with $g=4$ has degree $d $ possibly varying in the set $\{3,..., M-2\}$, and so in the following proof it is  tailored into two parts ($d\leq d_{max}$ and $d> d_{max}$) for evaluation. For the case of $d\leq d_{max}$, by comparing RIP-1 and RIP-3, we can derive that $\frac{3k-2}{4d+k-2}< \frac{(3k-2)s}{(k-2)s+4d}$, if $2\leq s \leq d-1$ and $3 \leq d \leq M/2$. This indicates that the RIC-$\delta_k$ of binary matrix with $g>4$ is less than that of binary matrix with $g=4$ under the same degree $d$.  So considering  $A(M,N,d_{max})$ is the best among all binary matrices with $g>4$, it also  outperforms the binary matrix with $g=4$ and $d\leq d_{max}$. For the case of $d> d_{max}$, by comparing RIP-1 and RIP-3,  it can be shown that $A(M,N,d_{max})$ performs better than the binary matrix with $g=4$ in the following two cases: \begin{itemize} \item [1)] $d_{max}\geq d/s$, derived with $\frac{3k-2}{4d_{max}+k-2}\leq \frac{(3k-2)s}{(k-2)s+4d}$ under $d_{max} <d \leq M/2$; \item [2)] $d_{max} \geq \frac{(k+1)(2d-M)}{6s+2(2d-M)}$, deduced from $\frac{3k-2}{4d_{max}+k-2}\leq\frac{(3k-2)s+(k-2)(M-2d)}{(k-2)s-(M-2d)k+2M}$ under  $M/2 <d \leq M-2$.\end{itemize}

 %Otherwise, under the inverse conditions, there possibly exists binary matrix with $g=4$ better than $A(N,M,d_{max})$, though up to now we have no idea whether the better matrix of above structure really exists or not. Due to this deficiency,  $A(N,M,d_{max})$ is assigned as 'near-optimal' rather than 'optimal'.

\end{IEEEproof}
%with RIP-1 or RIP-2, it is easy to derive that the RIC-$\delta_k$ of LDPC matrix  $A(N,M,d)$  decreases as  the degree $d$ increases. $\Box$ %If $d=d_{max}+1$, LDPC mtrix with $g=4$ beyond the \emph{Definition 1}, takes coherence $\mu=2/d >1/d_{max}$ values from ${0,1/d (<1/d_{max}), 2/d ~(>1/d_{max})} $ $\Box$

\vspace{3pt}
%\noindent{\emph{Remarks:}}
\textbf{Remark:} According to Theorem 4, we know that  there are  two reverse conditions  for which   the binary matrix with $g=4 $ will outperform the near-optimal matrix $A(M,N,d_{max})$. However, in practice   it seems hard to construct such kind of matrices  based on the observations below:
 \begin{itemize}
 \item For the reverse  condition of $d>sd_{max}$, it seems hard to construct the binary matrix with a desired degree $d$. Indeed, based on the definition of $d_{max}$, it is known that $d\leq sd_{max}$ if $s=1$.  Further, let $d'=d_{max}+1$, and with the definition of binary matrix with $g=4$, the corresponding $s'$ should take value from the set $ \{2,...,d'-1\}$ . In this case, it is easy to  derive that that $d'/s'<d_{max}$. As for $d'>d_{max}+1$, with Lemma 2, it can be observed that with high probability  $s'$ will  take larger values as $d'$ increases.  Empirically,  $s'$ often increases much faster than $d'$ during the practical matrix construction. Therefore it can be conjectured that $d'/s'<d_{max}$ when $d'>d_{max}+1$.
      %Further, let $s'$ be the smallest value that can be achieved by binary matrix with $g=4$ and $d=d'$, we can also deduce that $d''/s''<d'/s'$ where $d''=d'+1$ and $s''\in \{s'+1,...,d''-1\}$ with the law that $s'$ cannot decrease as $d'$ increases. By induction, it is proved that $d>sd_{max}$ is impossible.
 \item With the reverse condition of $d_{max} < \frac{(k+1)(2d-M)}{6s+2(2d-M)}$, it seems difficult to directly analyze or determine the binary matrix with desired  degree $d$. But  note that the reverse condition is derived on the assumption that $M/2 <d \leq M-2$, which is out of our interest of searching the sparsest matrix in the context that sparse matrix can propose comparable performance with dense matrix.
 \end{itemize}
%Consequently, considering practical application, it is reasonable to conjecture that $A(M,N,d_{max})$ probably possesses the 'near-optimal' sensing performance.
Besides above two reverse cases, in fact, there remains  a specific subclass of binary matrices with $g=4$, which possibly outperforms  $A(M,N,d_{max})$. This type of matrices has degree $d$ slightly larger than $d_{max}$, such that the associated bipartite graph tends to hold relative few shortest cycles with length equal to 4, and equivalently with high probability the nonzero correlations between distinct columns thus take value $1/d$ ($<1/d_{max}$) rather than $\mu=s/d$ ($>1/d_{max}$), where $s$ is also slightly larger than 1. In this case, this type of matrices can be approximately regarded as the binary matrices with $g>4$ but $d$ ($>d_{max}$), and so they probably obtain better RIP than $A(M,N,d_{max})$.   In practice, this type of matrices tends to occur at a  relative small region, e.g.,  $d-d_{max}<3$ in our experiments, since with the observation on Lemma 2, the probability of taking correlation value $1/d$ will dramatically decrease as  $d$ increases. This  means that their performance gain  over $A(M,N,d_{max})$ is relatively small, as can be seen from the following experiments. In addition, it  is interesting to understand why this specific case is not disclosed in the proof of Theorem 4. As the remark of Theorem 3, this is because  the RIP of the matrices with high probability taking correlation values $1/d$ rather than $\{2/d,...,s/d\}$, cannot be accurately described with RIP-3, such that they are ignored during the RIP comparison of Theorem 4.

%It  is interesting to understand why this specific case is not disclosed in the proof of Theorem 4.  This is because currently  RIP can only be approximated with some extreme conditions, and so in fact it cannot perfectly characterize all possible  matrices.  Exactly speaking, in this paper the bounds of  RIP can only be achieved as the correlation between columns takes value $\mu=s/d$ with high probability \cite{zhan06}, while this condition obviously cannot be well satisfied by the specific matrices mentioned above, which take correlation values $1/d$ rather than $s/d ~(>1/d)$ with high probability.

Note that in this paper the binary matrix is evaluated only with the \emph{regular} form. Similar conclusion  can  be  expanded to  the \emph{irregular} binary matrix of uneven degrees, that is,  the irregular matrix with larger average degree  tends to have better RIP when $g>4$. Significantly, in practice the irregular matrix  probably obtains better RIP than the regular matrix, since the former usually can be constructed with larger average degree than the latter under the constraint of $g>4$ \cite{Huxiaoyu05}.

\section{Simulation results}

\subsection{Simulation setup}

The proposed near-optimal binary sensing matrix $A(N,M,d_{max})$ is evaluated by comparing it with four types of matrices below:\begin{itemize}
                                                                                                                            \item [1)] deterministic binary matrix with $g>4$ and $d<d_{max}$ constructed with PEG algorithm;
                                                                                                                            \item [2)] deterministic  binary matrix with $g=4$ and $d>d_{max}$ constructed with PEG algorithm;
                                                                                                                            \item [3)] random binary matrix with uniform column degree;
                                                                                                                            \item [4)] random Gaussian matrix.
                                                                                                                          \end{itemize}
  Recall that up to now the binary matrix with $g=4$ and $\mu=s/d$ cannot be explicitly constructed. So here we exploit two typical binary matrices with $g=4$ while without specific constraint on $\mu$: the binary matrix with $g=4$ and $d>d_{max}$ constructed with PEG algorithm, and random binary matrix $R(M,N,d)$. For notational clarity, in the following part all binary matrices (with $g\geq 4$) constructed with PEG algorithm are denoted with $A(M, N, d)$. Gaussian matrix is referred as a  performance baseline. Here PEG  algorithm is adopted to construct the binary matrices with $g>4$  for the following two reasons. First, this greedy algorithm based on maximizing girth is suitable for approaching the $d_{max}$ of binary matrix  with $g>4$ in terms of the fact that the girth $g$ will decrease dramatically as $d$ increases. Second, it can flexibly construct binary matrices with diverse degrees. However,  due to the greediness, it should be noted that PEG algorithm cannot guarantee to obtain the theoretical $d_{max}$ in practice.  For limited simulation time, here we only test the matrices of size (200, 400). For other matrix sizes, the interested readers may refer to the experimental results shown in \cite{Weizhi12}. Given the matrix size of (200, 400),  the near-optimal matrix with $d_{max}=7$, namely $A(200,400,7)$, is  determined with PEG algorithm.

The simulation exploits four representative decoding algorithms: orthogonal matching pursuit (OMP) algorithm \cite{Pati93} \cite{Tropp}, iterative hard thresholding (IHT) algorithm \cite{Blumensath09}, subspace pursuit (SP) algorithm \cite{Dai09} and basis pursuit (BP) algorithm \cite{Boyd04}. Each simulation point is derived after $10^4$ iterations. Both binary random matrix and Gaussian matrix are randomly generated at each iteration. The  sparse signal has nonzero entries  drawn from $N(0,1)$. And the correct recovery rates are measured with $1-||\hat{x}-x||_2/||x||_2$.

\subsection{Near-optimal performance over varying sparsity}

\begin{center}
\begin{table}[t]
\centering

\caption{The largest sparsity level $k$  that can be recovered with probability larger than $99\%$, for four classes of matrices:  $A(200,400,d\leq 7)$ with $g>4$ (namely $A_\ell$), $A(200,400,d> 7)$ with $g=4$ (namely $A_e$), random binary matrix $R(200,400,d)$ (namely $R$) and Gaussian random matrix of size (200,400) (namely $G$).  The largest $k$ over varying $d$ is highlighted in bold.  Recall that $A_\ell$ with $d=7$ denotes the proposed near-optimal matrix $A(200,400, 7)$.}
%\vspace{4pt}
{\footnotesize%\scriptsize
\begin{tabular}{|c|c|c|c|c|c|c|c|c|c|c|c|c|c|c|c|c|c|c|c|c|c|c|}

\hline
\multicolumn {3}{|c|} {$d$} &2&3&4&5&6&\textbf{7}&8&9& 10& 11& 12& 13& 14& 15& 20& 30& 40& 50& 100\\
\hline

\hline
\hline
\multirow{4}{*}{\rotatebox{90}{OMP} }&\multirow{4}{*}{$k$ }&$A_\ell$  & 29&   70&   75  &  78 & 80  &$\mathbf{81}$ & -  & \ -  &  -  &  - &   - &  - &  -&   - &   -  &  -&   - &  -&   -\\
&&$A_e$  & -&   -&    -  &   - &  -  & - & \textbf{83}  & 83  & 81   & 80 &  79 & 78 & 78&  77 &  75  & 74&  48 & 26&    2\\
&&$R$ &   0 &  55  &  69 &   73&  75 &  \textbf{76} &  76 &  76 & 76  &  76 & 76  & 76  &76 & 76 & 76 & 76  & 76 & 76  & 76\\\cline{4-22}
&&$G$ &  \multicolumn {19}{|c|} {76}\\
\hline

\hline
\hline
\multirow{4}{*}{\rotatebox{90}{IHT} }&\multirow{4}{*}{$k$ }&$A_\ell$  & 1&   34&   47  &  53 & 55  & \textbf{56 } & - &- & - & -  & - & -& - & -& -& -  & -& - & -\\
&&$A_e$  &-&   -&   -  &  - & -  & - & \textbf{57} &55 & 53  & 50   & 48 & 47 & 45 & 44&  37&  27  & 16&  7 & 1\\

&&$R$&    0  &  14 &   38& 45 &  \textbf{48} &  48 &  47 & 46  &   45  & 44  &44 & 44 & 43& 43 & 38 & 30  & 22&18&3\\\cline{4-22}
&&$G$ &  \multicolumn {19}{|c|} {53}\\
\hline

\hline
\hline
\multirow{4}{*}{\rotatebox{90}{SP} }&\multirow{4}{*}{$k$ }&$A_\ell$  &17&   62&  71  &  73 & 74  & \textbf{75}  & -  &  -   &  -    &  - &  -  &  -  &  - &   -  &   -   &  - &   -  & - &    - \\
&&$A_e$  & - &   - &   -   &   -  &  -   & -   & \textbf{75}  & 74  & 74   & 73 &  72 & 71 & 71&  77 &  71  & 70&  25 & 9&    1\\
&&$R$ &   0 &  48  &  65 &   68&  \textbf{71}&  71 &   71 &  71 &  71  &   71 &  71  &  71  &70 &70  & 70  &70  & 70  & 70   & 69\\\cline{4-22}
&&$G$ &  \multicolumn {19}{|c|} {73}\\
\hline

\hline
\hline
\multirow{4}{*}{\rotatebox{90}{BP} }&\multirow{4}{*}{$k$ }&$A_\ell$  & 25&   57&   \textbf{61} &  61 & 61  & 61  & - &- & -  & -   & - & - & - & -&  - &  -  & -& - & -\\
&&$A_e$  & -&   -&   -  &  - & -  & -  & \textbf{62} &61 & 59  & 59   & 58 & 58 & 57 & 57&  54 &  51  & 36&  18 & 1\\
&&$R$&    0  &  45 &   55&  \textbf{58} &  58 &  58 &  58 & 58  &   57  & 57  &57 & 56 & 56& 56  & 55 & 53  & 50&49&37\\\cline{4-22}
&&$G$ &  \multicolumn {19}{|c|} {63}\\
\hline

\end{tabular}
}
\end{table}
\end{center}

The binary matrices of varying degree $d$ are evaluated by the maximum sparsity $k$ of sparse signal that can be  correctly recovered with a rate over $99\%$.   Obviously, larger $k$ indicates better performance. As performance reference, the maximum $k$ for Gaussian matrix is also provided. All results are shown in Table 1. For notational simplicity, in Table 1 the binary matrices $A(200,400,d)$ constructed with PEG algorithm are shortly denoted as $A_\ell$ and $A_e$ respectively for the cases $g>4$ and $g=4$; and random binary matrix $R(200,400,d)$ and Gaussian matrix are abbreviated to $R$ and $G$, respectively.  Note that, for limited simulation time, we cannot enumerate all possible values of $d$. But clearly the results are sufficient to capture the performance varying tendency of tested matrices.  With these results, first, it can be observed that the maximum $k$ follows the order: the near-optimal matrix $A(200,400,7)$ $>$Gaussian matrix$>$Random binary matrix $R(200,400,2\leq d\leq 100)$, except the unique case of Gaussian matrix$>$the near-optimal matrix$>$Random matrix under BP decoding. This demonstrates that the near-optimal matrix $A(200,400,7)$ outperforms random binary matrix. And then we turn to compare the near-optimal matrix $A(200,400,7)$ with other binary matrices $A(200,400,d)$ constructed with PEG. Among all binary matrices of  $g>4$ (namely $A_\ell$ in the Table 1),  clearly  $A(200, 400,7)$ is indeed the only case that  can achieve the best performance simultaneously for above four decoding algorithms. Note that, although the matrices $A(200, 400,d\in \{4, 5, 6\})$ achieve same $k$ with $A(200, 400, 7)$ under BP decoding, their correct decoding precisions in fact are less than the latter. However, compared with the cases of $g=4$ (namely $A_e$ in the Table 1),  there are few cases   obtaining comparable and even better performance than the proposed near-optimal case, such as $A(200, 400,d\in \{8 , 9\}>d_{max}=7)$ under OMP and $A(200, 400,d=8)$ under other three algorithms.  With the former remarks of Theorem 4, these results can be explained by the fact that the aforementioned matrices $A(200, 400,d\in \{8 , 9\})$ constructed with PEG,  with high probability take nonzero correlation values as $1/d$ ($<1/d_{max}$) rather than as $\mu=s/d$ ($>1/d_{max}$), if $d-d_{max}$ is relatively small, so that they can be approximately regarded as  the binary matrices with $g>4$ but $d>d_{max}$. Recall that PEG algorithm is designed to greedily reduce the increasing speed of the girth of bipartite graph as the degree $d$ of the binary matrix progressively increases. This yields that the correlation values  of the constructed matrix largely center on $1/d$ rather than on $s/d$ with $s$ slightly larger than  1, when $d$ is slightly larger than $d_{max}$. With the results shown in Table 1 and \cite{Weizhi12}, it is obvious that this type of  matrices constructed with PEG algorithm lies in a relative small region, e.g. $d-d_{max}\leq 2$ in our simulations. Therefore they practically can be easily derived after the near-optimal binary matrix is determined. Overall, the proposed binary matrix indeed shows  nearly the best performance with  the highest sparsity.

 Moreover, it is interesting to point out that the binary matrices constructed with PEG, $A(200,400,d<d_{max})$, still outperform random binary matrix and even Gaussian matrix for most decoding algorithms, if $d$ is slightly smaller than $d_{max}$, e.g. $d_{max}-d<3$ in our experiments. This allows us to practically construct the binary matrix with a more hardware-friendly  structure \cite{LiuXinji13}, i.e. the quasi-cyclic structure, while  preserving favorable performance in the negative case where the quasi-cyclic structure tends to slightly lower the value of $d_{max}$ \cite{Zongwang04} \cite{LiuXinji13}.

%Binary matrices of varying degrees are evaluated by the maximum sparsity $k$ of sparse signal that can be  correctly recovered with rate over $99\%$.   Obviously, larger $k$ indicates better performance.  The results are shown Table 1. As reference, the maximum $k$ for Gaussian matrix is also derived as 76 under OMP decoding and 63 under BP. As it is expected, the near-optimal binary matrix, $A(200, 400,d_{max}=7)$, achieves nearly the best performance with $k=81$, larger than $k=76$ for both binary random matrix. In contrast, it performs  slightly worse than two binary matrices $A(200, 400,d\in \{8 , 9\})$ with $k=83$. With the former remarks for Theorem 4, this can be explained by the fact that two matrices $A(200, 400,d\in \{8 , 9\})$ constructed with PEG algorithm take nonzero correlation values as $1/d$ ($<1/d_{max}$)with higher probability rather than as $\mu=s/d$ ($>1/d_{max}$), $s>1$. Recall that PEG algorithm is designed to greedily reduce the increasing speed of girth of bipartite graph as degree $d$ of binary matrix progressively increases. This yields that the correlation values  of constructed matrix largely center on $1/d$ rather than $s/d$ of $s$ slightly larger than  1, when $d$ is slightly larger than $d_{max}$. With the results shown in Table 1 and \cite{Weizhi12}, it is obvious that this type of  matrices constructed with PEG algorithm lies in a relative small region of $d-d_{max}<2$, thereby  practically  they can be easily derived after the near-optimal binary matrix is determined.

\subsection{Performance over sparse signals of low sparsity or Gaussian noise}

%Considering in real applications the sparsity of sparse signal is likely to be uncontrollable   and the noise  is also inevitable, this section  evaluates the proposed near-optimal matrix $A(200,400,7)$ from above two aspects. Random binary matrix $R(200,400,7)$ \footnote{ Note that random binary matrix  has achieved its best performance at $d=7$ as shown in Table 1.} and Gaussian matrix are also tested for comparison. The performances over sparse signal of varying sparsity $k$ are illustrated in Figure 2. Similar with the results shown in Table 1, here the near-optimal matrix performs better than random binary matrix, and slightly worse than Gaussian matrix under BP.  In Figure 3, we depicts the influences of Gaussian noise  $N(0,\sigma ^2)$ on \emph{normalized} sparse signal of $k=40$. The near-optimal matrix outperforms  other two competitors with  obvious gain under OMP and tiny gain under BP.  In short, the near-optimal matrix obtains better overall performances over random binary matrix and even Gaussian matrix.

 %obtain obvious performance gains over  This means that in practice the prosed matrix is allowed to replace the classical Gaussian matrix with not only significant complexity reduction but also performance gain.

\begin{figure}[t]
\centering
\graphicspath{fig}
\includegraphics[width=0.495\textwidth]{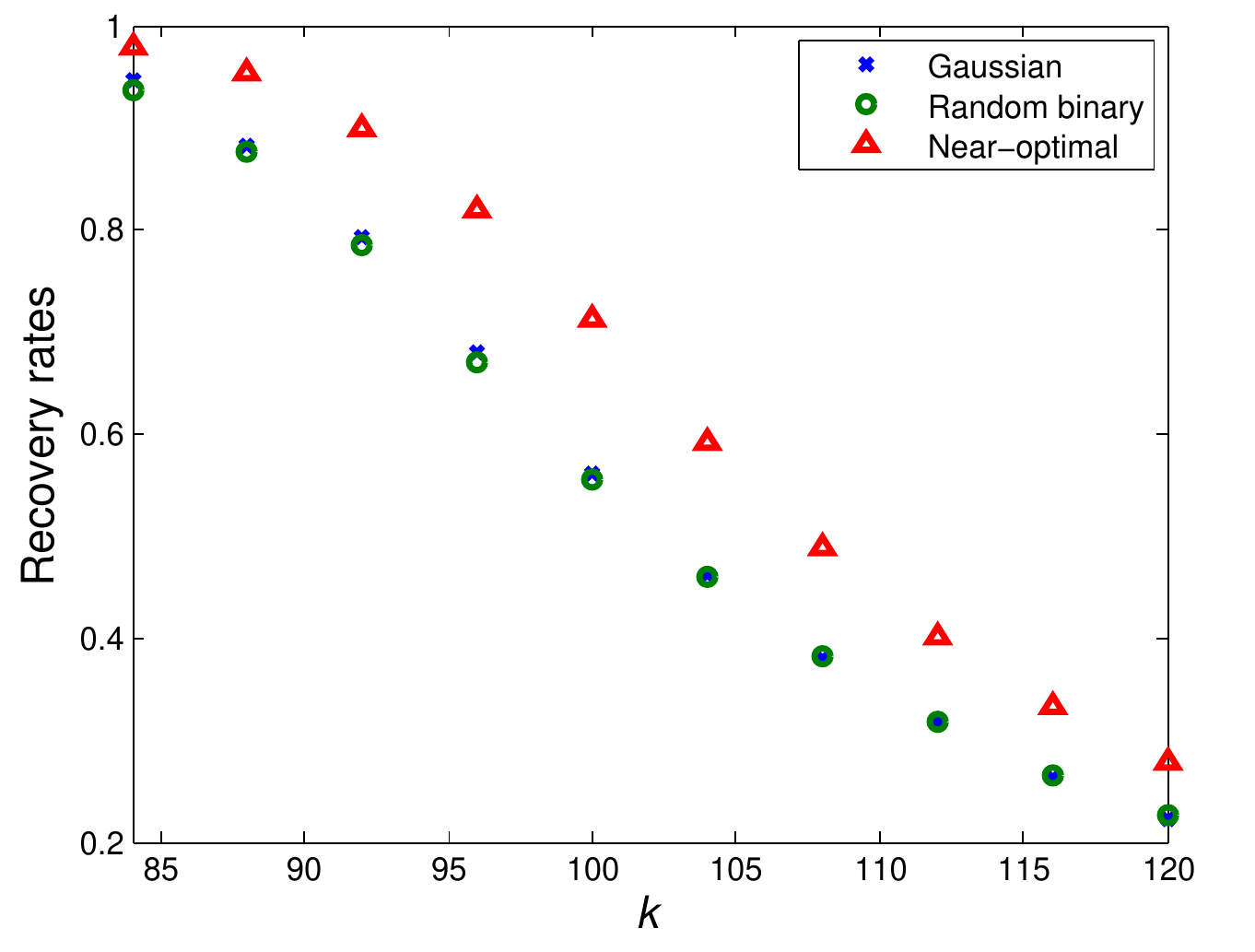}
\includegraphics[width=0.495\textwidth]{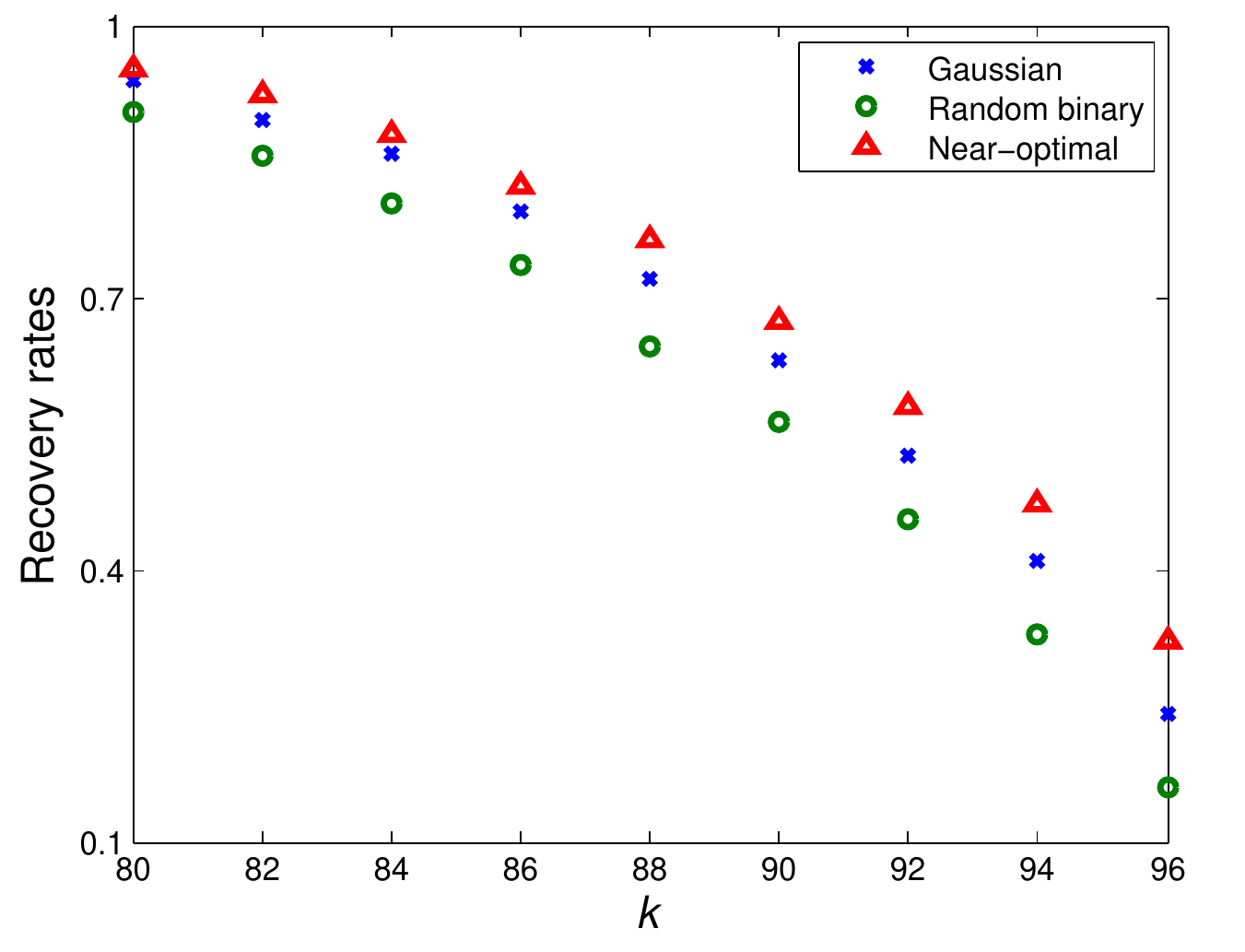}\\
(a)~~~~~~~~~~~~~~~~~~~~~~~~~~~~~~~~~~~~~~~~~~~~~~~~~~~~~~~~~~~~~(b)\\
\vspace{6pt}
\includegraphics[width=0.495\textwidth]{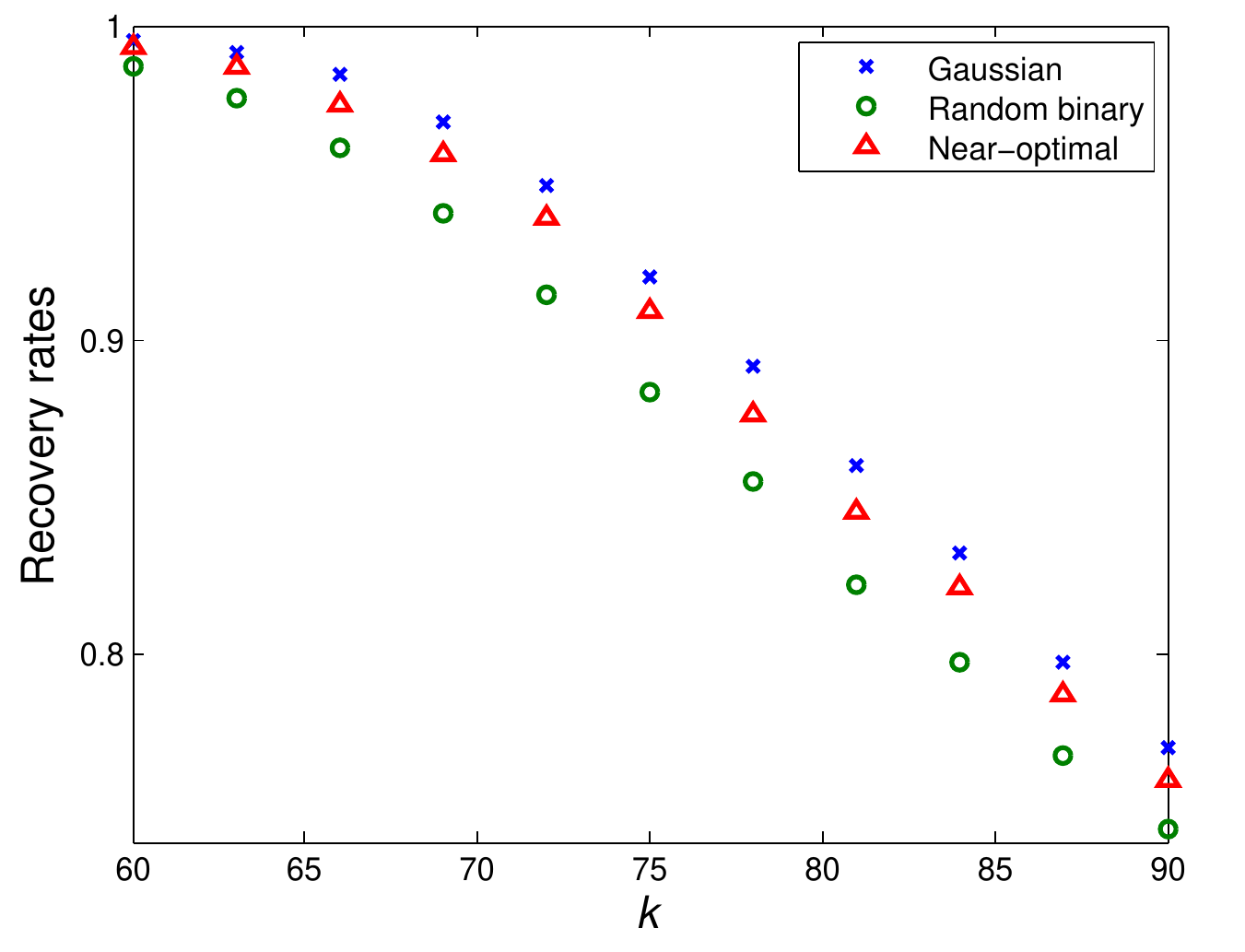}
\includegraphics[width=0.495\textwidth]{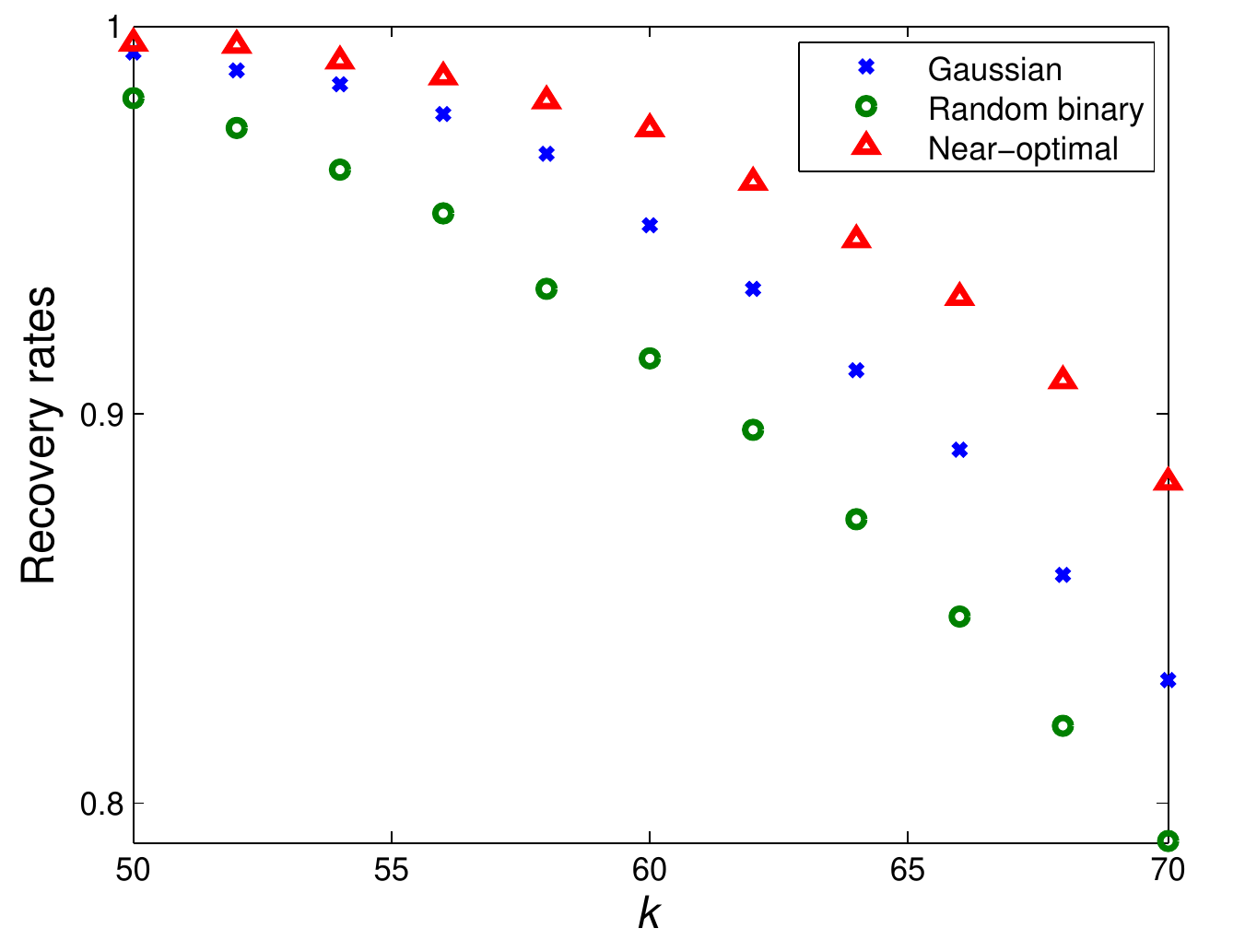}\\
(c)~~~~~~~~~~~~~~~~~~~~~~~~~~~~~~~~~~~~~~~~~~~~~~~~~~~~~~~~~~~~~(d)
\caption{The  recovery rates of the near-optimal matrix $A(200,400,7)$,  random binary matrix $R(200,400,7)$ and Gaussian matrix,  over sparse signals of varying sparsity  $k$. OMP in (a), SP in (b), BP in (c) and IHT in (d).}
\end{figure}

\begin{figure}[!]
\centering
\graphicspath{fig}
\includegraphics[width=0.495\textwidth]{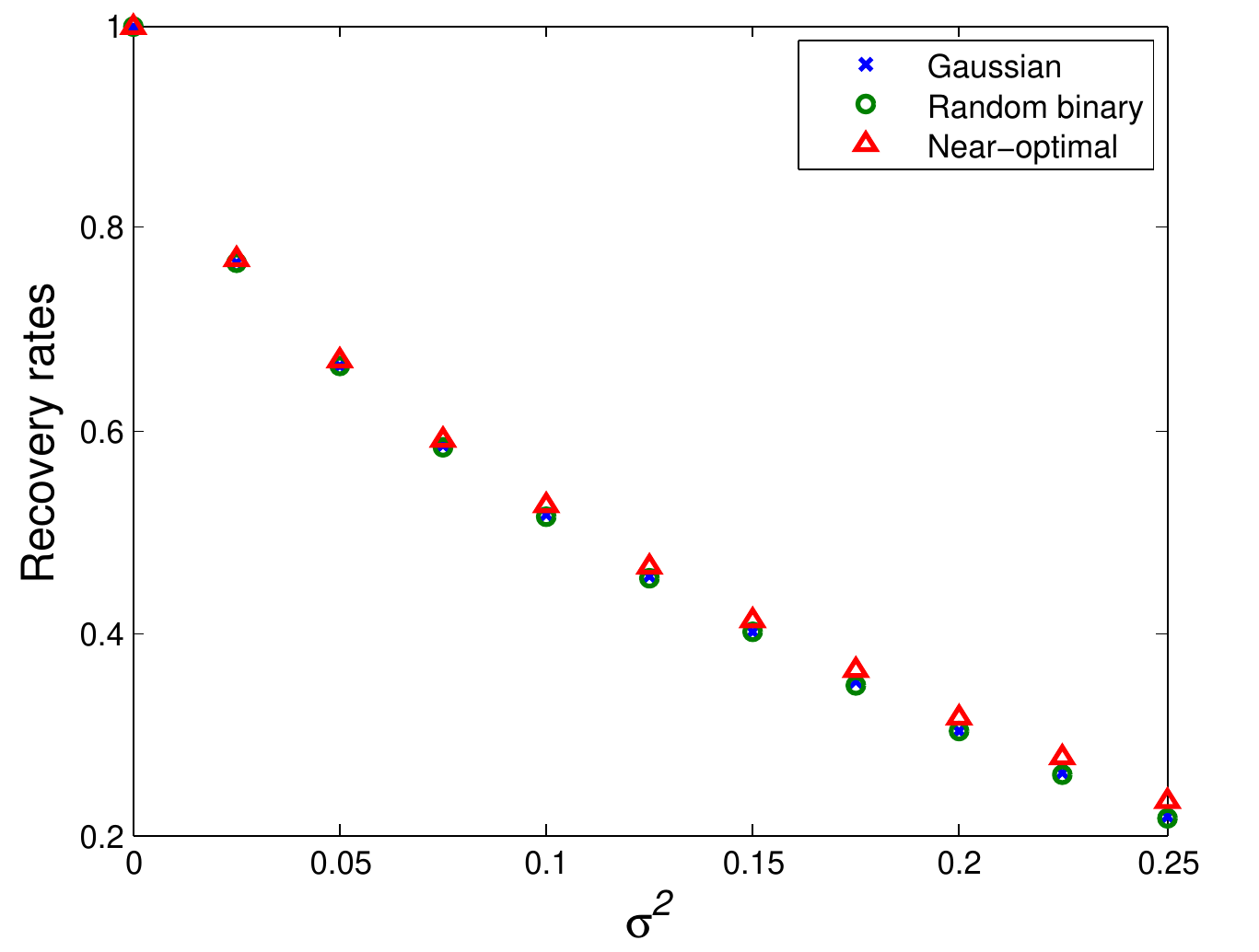}
\includegraphics[width=0.495\textwidth]{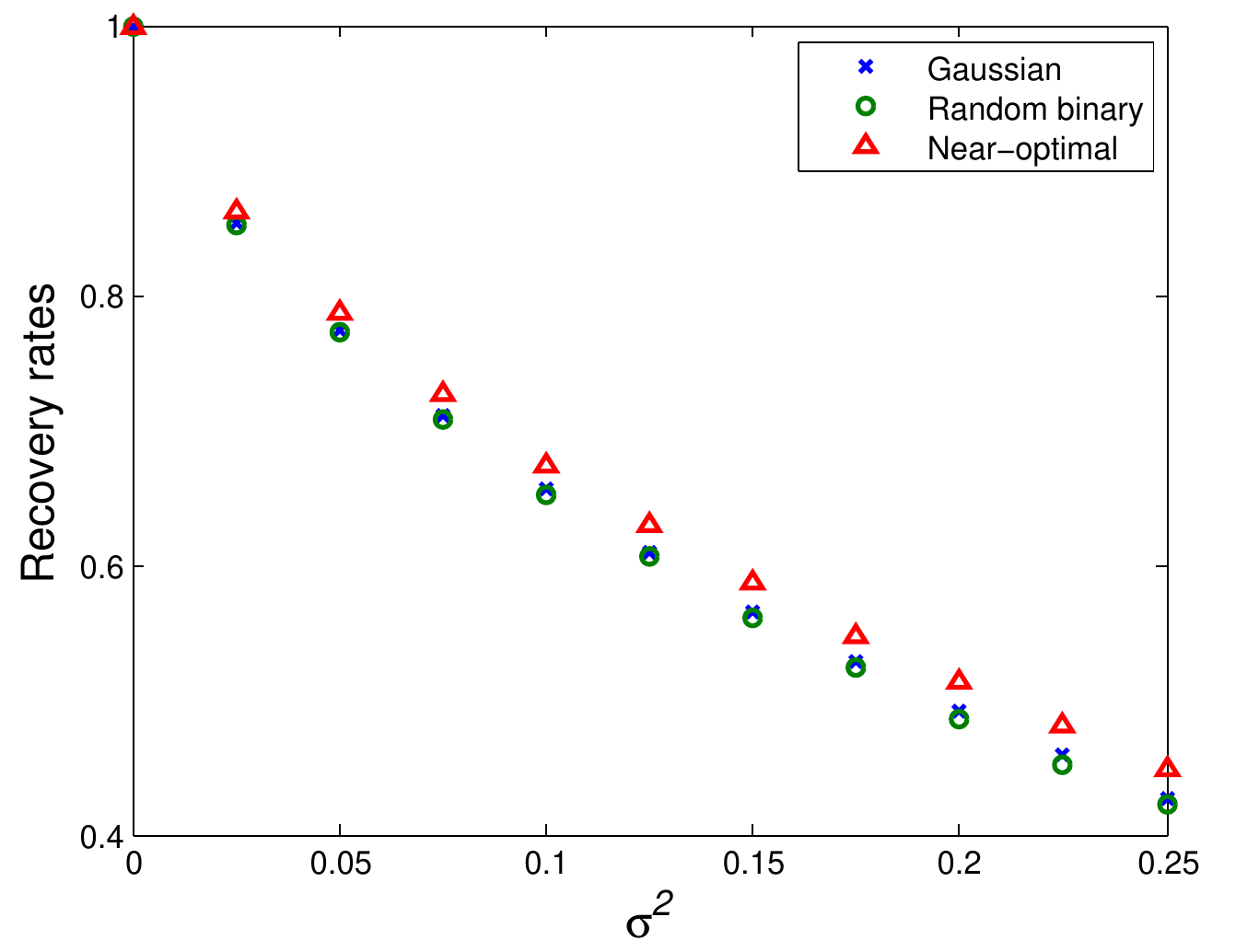}\\
(a)~~~~~~~~~~~~~~~~~~~~~~~~~~~~~~~~~~~~~~~~~~~~~~~~~~~~~~~~~~~~~(b)\\
\includegraphics[width=0.495\textwidth]{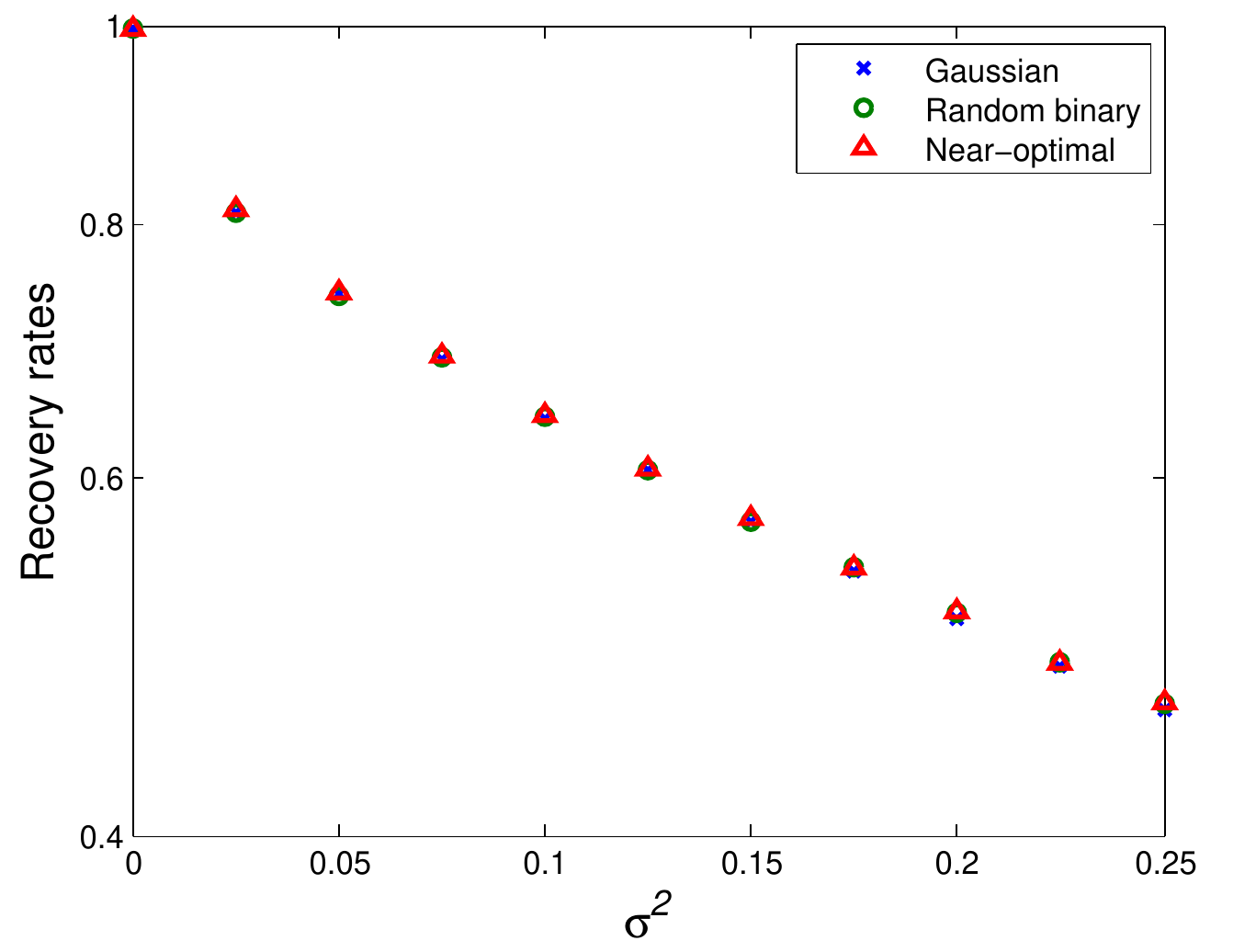}
\includegraphics[width=0.495\textwidth]{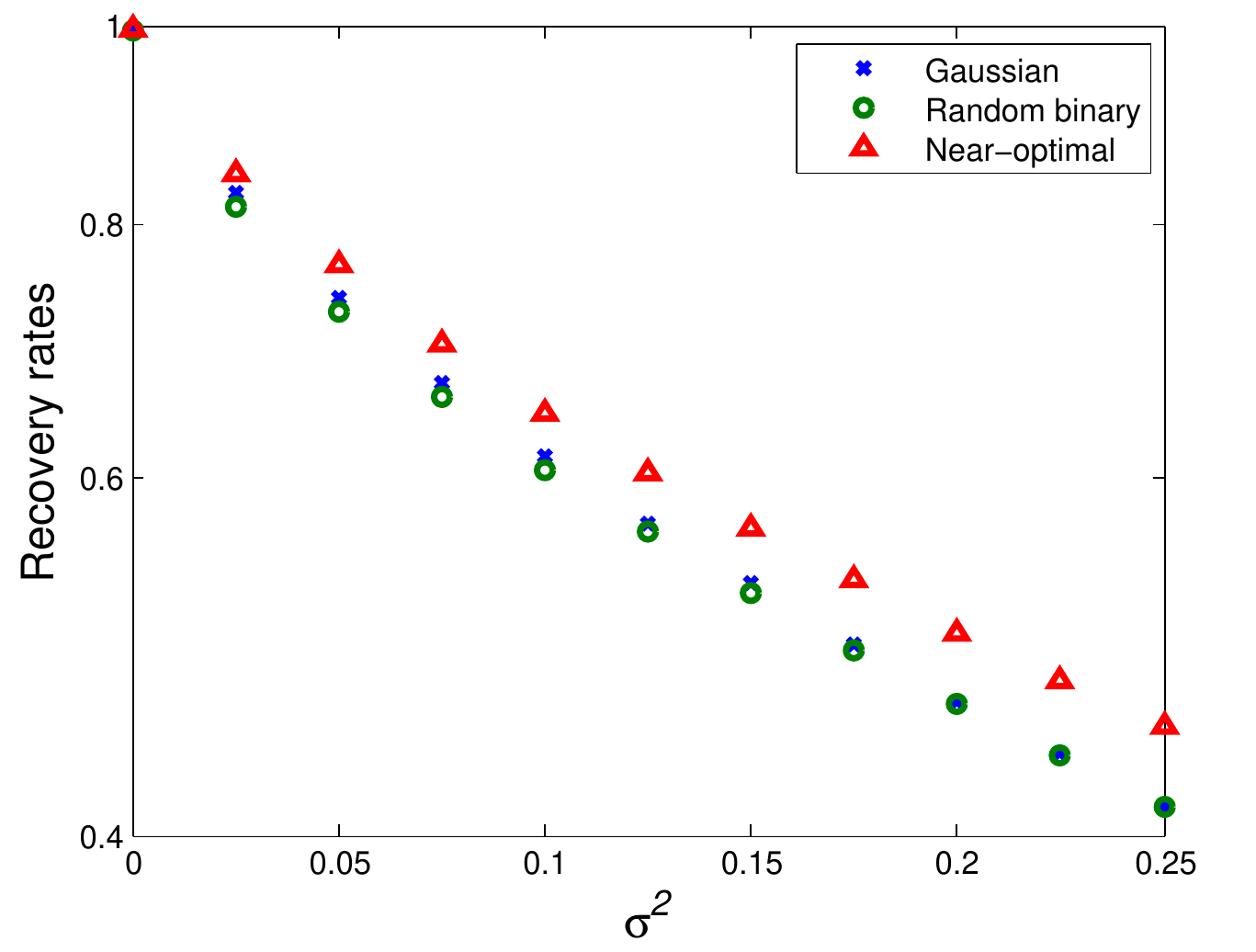}\\
(c)~~~~~~~~~~~~~~~~~~~~~~~~~~~~~~~~~~~~~~~~~~~~~~~~~~~~~~~~~~~~~(d)
\caption{The  recovery rates of the near-optimal binary matrix $A(200,400,7)$, random binary matrix $R(200,400,7)$ and Gaussian matrix, over normalized sparse signals perturbed with Gaussian noise $N(0,\sigma^2)$. OMP  in (a), SP in (b), BP in(c) and IHT in(d).}
\end{figure}

This section evaluates the practical performance of the near-optimal matrix  with sparse signal suffering from the following two potential challenges: 1)  sparsity $k$ beyond the tolerance limit of sensing matrix; 2) additive Gaussian noise. Random binary matrix $R(200,400,7)$\footnote{ Note that random binary matrix  has achieved its best performance at $d=7$ for  above four decoding algorithms as shown in Table 1.} and Gaussian matrix are also tested for comparison. The performance over sparse signals of  excessive sparsity $k$ is illustrated in Figure 3.  In Figure 4, we depict the influence of Gaussian noise  $N(0,\sigma ^2)$ on \emph{normalized} sparse signals of the sparsity $k=40$, which can be well decoded by  three types of matrices as shown in Table 1, such that the following comparison under noises is fair. Similar with the results shown in Table 1, the proposed near-optimal matrix still shows  better  performance than other two types of matrices, except for the case of sparse signals of excessive $k$ with BP decoding, as shown in Figure 3(c), where it performs slightly worse than Gaussian matrix. In addition, due to the low performance resolution of  Figure 4(c),  it is necessary to point out that the near-optimal matrix also obtains tiny  gains over other two competitors on the case of sparse signals of Gaussian noise decoded by BP.
\section {Conclusion}

This paper has proposed the near-optimal distribution of binary sensing matrices through the analysis of  RIP. In practice, the proposed matrix of expected performance can be approximately constructed with PEG algorithm. Specifically, it even shows better performance over Gaussian matrix with  popular greedy decoding algorithms. As stated before, the term 'near-optimal' is derived due to the fact that in practice there exists a class of matrices with  sightly better RIP.  These  matrices hold degrees slightly larger than that of the near-optimal matrix, such that they can be easily found in practice. However, they are not formally defined in the literature since  their structures are hard to be explicitly formulated. One must note that, as a sufficient condition, RIP is not an ideal tool for evaluating the performance of sensing matrices. So a more effective way is expected to be developed  in the future to  tackle this problem. In addition, it should be mentioned that the ideal degree of the proposed near-optimal matrix is only approximately bounded in this paper; and the  practical construction algorithm, PEG algorithm, is also suboptimal due to its greediness. Consequently, it might be interesting in the future to further investigate the real degree of the proposed near-optimal matrix both in theory and practice.

\appendices
\section{Proof of Theorem 1}
\begin{IEEEproof}  As stated before, the solution to RIC-$\delta_k$ can be reformulated as the pursuit for the extreme eigenvalues of random symmetric matrix $A_T'A_T\in \{0,1,1/d\}^{k\times k}$, where $|T|=k$. Thus the following proof borrows the solution algorithm of extreme eigenvalues proposed in \cite{zhan06}. The eigenvalues of $A_T'A_T$ are customarily denoted and ordered with  $\lambda_1(A_T'A_T)\geq\ldots\geq\lambda_k(A_T'A_T)$ .
\begin{enumerate}

\item Let $B=A'_TA_T-I\in \{0,1/d\}^{k\times k}$,  then $B_{ii}=0$ and $B_{ij,i\neq j} =0$ or $1/d$.
\\
Let normalized $x=(x_1,\ldots,x_k)'$ be the eigenvector corresponding to $\lambda_k(B)$. Then the minimal eigenvalue can be formulated as
$$\lambda_k(B)=x'Bx=\mathds{1}'[B\circ(xx')]\mathds{1}$$
where $\circ$ denotes the Hadamard product and $\mathds{1}=(1,\ldots,1)'\in\mathbb{R}^k$. Since $B$ is symmetric, by simultaneous permutations of the rows and columns of $B$, we can suppose  $x_i\geq 0$ for $i=1,\ldots,n$ and $x_i<0$ for $i=n+1,\ldots,k$, and then $xx'$ is divided into four parts:
\\
$$xx'=
\begin{bmatrix}
X_{n\times n}&X_{n\times (k-n)}\\
X_{(k-n)\times n}&X_{(k-n)\times(k-n)}
\end{bmatrix}
$$
where the entries in $X_{n\times n}$ and $X_{(k-n)\times(k-n)}$ are nonnegative, while the entries in $X_{n\times (k-n)}$ and $X_{(k-n)\times n}$ are nonpositive.   Further, define a novel matrix $\tilde{B}$ of same size with $B$
$$
\tilde{B}=
\begin{bmatrix}
0\times\mathds{1}_{n\times n}&\frac{1}{d}\times\mathds{1}_{n\times (k-n)}\\
\frac{1}{d}\times\mathds{1}_{(k-n)\times n}&0\times\mathds{1}_{(k-n)\times(k-n)}
\end{bmatrix}
$$
where $\mathds{1}_{a\times b}$ is an $a\times b$ matrix with all entries equal to $1$. It is easy to deduce that
$$
\lambda_k(\tilde{B})=\min\{y'\tilde{B}y:\|y\|=1\} \leq x'\tilde{B}x\leq x'Bx=\lambda_k(B).
$$
Since the rank of $\tilde{B}$ is at most $2$, it has at most two nonzero eigenvalues. Considering the trace and the Frobenius norm, we have
$$\lambda_k(\tilde{B})=-\sqrt{\frac{n(k-n)}{d^2}}, ~0\leq n\leq k.$$
If $k$ is even, $\lambda_k(\tilde{B})\geq-\frac{k}{2d}$, with '$=$'  at $n=k/2$.
\\
If $k$ is odd, $\lambda_k(\tilde{B})\geq-\frac{\sqrt{k^2-1}}{2d}$, with '$=$' at $n=(k-1)/2$ or $n=(k+1)/2$. %For $-\frac{\sqrt{k^2-1}}{2d}>-\frac{k}{2d}$, then $\lambda_k(\tilde{B})\geq-\frac{k}{2d}$.
\\
Then $\lambda_k(B)\geq \lambda_k(\tilde{B}) \geq-\frac{k}{2d}$, with the limitation  attained at $k$ is even and $n=k/2$.

So, we have the minimum eigenvalue $\lambda_k(A_T'A_T)\geq1-\frac{k}{2d}$.

\item Let $C=A'_TA_T-\frac{d-1}{d}\times I$, then $C_{ii}=1/d$ and $C_{ij,i\neq j } =0$ or $1/d$ .

Let normalized $x=(x_1,\ldots,x_k)'$ be the eigenvector corresponding to $\lambda_1(C)$. By simultaneous permutations of $C$ and $x$,  we can suppose  $x_i\geq 0$ for $i=1,\ldots,n$ and $x_i<0$ for $i=n+1,\ldots,k$, and the maximal eigenvalue is formulated as
$$\lambda_1(C)=x'Cx=\mathds{1}'[C\circ(xx')]\mathds{1}.$$
Further define
$$
\tilde{C}=
\begin{bmatrix}
\frac{1}{d}\times\mathds{1}_{n\times n}&0\times\mathds{1}_{n\times (k-n)}\\
0\times\mathds{1}_{(k-n)\times n}&\frac{1}{d}\times\mathds{1}_{(k-n)\times(k-n)},
\end{bmatrix}
$$
then
\[
\begin{split}
\lambda_1(\tilde{C})&=\max\{y'\tilde{C}y:\|y\|=1\}\geq x'\tilde{C}x\geq x'Cx\\&=\lambda_1(C).
\end{split}
\]
Since the rank of $\tilde{C}$ is at most $2$, it has at most two nonzero eigenvalues. Considering the trace and the Frobenius norm, we have
$$\lambda_1(\tilde{C})=\frac{k+|k-2n|}{2d}.$$
Then $\lambda_1(C)\leq\lambda_1(\tilde{C})\leq\frac{k}{d}$, with '$=$' at $n=0$ or $n=k$.
Thus, we can further derive  $$\lambda_1(A_T'A_T)=\lambda_1(C)+\frac{d-1}{d}\leq\frac{k+d-1}{d}.$$

\item Finally,  it follows from the results of both 1) and 2) that $$\delta_k=\frac{\lambda_1(A'_TA_T)-\lambda_k(A'_TA_T)}{\lambda_1(A'_TA_T)+\lambda_k(A'_TA_T)}=\frac{3k-2}{4d+k-2},$$  with $\frac{\lambda_1(A'_TA_T)}{\lambda_k(A'_TA_T)}=\frac{1+\delta_k}{1-\delta_k}$ \cite{Foucart09}.
\end{enumerate}
\end{IEEEproof}
\section{Proof of Theorem 2}
\begin{IEEEproof}
To derive the extreme eigenvalues of $A_T'A_T$, we first search the extreme eigenvalues of $$B=(A_T'A_T-I)$$
where $I$ is an identity matrix. And clearly $B$ is a symmetric matrix of the diagonal elements equal to 0, and  the off-diagonal elements equal to $1/d$ with property $\rho$ and 0 with property $1-\rho$.

With \cite{Tran13}, suppose

$$Q=\frac{1}{\sqrt{\rho(1-\rho)}}(d B-\rho\mathds{1})$$
where $\mathds{1}$ is a all-ones matrix.  Then $Q$ has entries with mean zero and variance one. With \emph{Wigner semicircle law} \cite{Pastur72}, the extreme eigenvalues $\frac{1}{\sqrt {k}}Q$  with $k=|T|$, can be approximated as
$$
-2\leq \lambda(\frac{1}{\sqrt {k}}Q)\leq 2
$$
namely,
%$$
%-2{\sqrt {k}}\leq \lambda(Q)\leq 2{\sqrt {k}}
%$$
%and
$$
-2{\sqrt {k\rho(1-\rho)}}\leq \lambda (dB-\rho\mathds{1})\leq 2{\sqrt {k\rho(1-\rho)}},
$$
if $k\rightarrow \infty$ \cite{Furedi81}.

With \emph{cauchy interlacing inequality}  \cite{Tao11}, one can further derive
$$\lambda_i(dB-\rho \mathds{1}) \leq \lambda_i(dB) \leq \lambda_{i-1}(dB-\rho \mathds{1})$$
for $1 <i \leq k$, if $B-\rho \mathds{1}$ and $\rho \mathds{1}$ are Hermitian matrices, and $\rho \mathds{1}$  is positive semi-definite and has rank equal to 1.
As a result,  it is easy to derive that

$$\lambda_2(B)\leq \frac{1}{d}\cdot \lambda_1(dB-\rho\mathds{1})\leq  \frac{2}{d}{\sqrt {k\rho(1-\rho)}}$$
and
$$\lambda_k(B)\geq \frac{1}{d}\cdot \lambda_k(dB-\rho\mathds{1})\geq  - \frac{2}{d}{\sqrt {k\rho(1-\rho)}}$$

As for $\lambda_1(B)$ \footnote{In \cite{ando11}, it is proved that $\lambda_1(B)\approx k\rho$, as $k\rho$ is sufficiently large.}, it is known that \cite{yosh12}
$$\lambda_1(B)\approx \frac{1}{d}(k\rho+1).$$
In this sense, the extreme eigenvalues of $A_T'A_T$ can be approximately formulated as

$$\lambda_1 (A_T'A_T)=\lambda_1 {B}+1\leq \frac{1}{d}(k\rho+1) +1$$
and
$$\lambda_k (A_T'A_T)=\lambda_k {B}+1\geq - \frac{2}{d}{\sqrt {k\rho(1-\rho)}}+1$$
Finally, the RIC of $A_T'A_T$ is deduced as

$$\delta_k=\frac{\lambda_1-\lambda_k}{\lambda_1+\lambda_k}=\frac{k\rho+2\sqrt {k\rho(1-\rho)}+1}{k\rho-2\sqrt {k\rho(1-\rho)}+2d+1}$$
\end{IEEEproof}
\section{Proof of Theorem 3}
The proof is similar to that for Theorem 1 in Appendix A. So in the following we just give a sketch.
\begin{IEEEproof}
%Let $\lambda_1(*)$ and $\lambda_k(*)$ denote maximal and minimal eigenvalues, respectively.
 \begin{enumerate}
   \item  If $3\leq d\leq M/2$, $[A'_TA_T]_{ii}=1$ and $[A'_TA_T]_{ij,i\neq j}\in \{0,\ldots,s/d\}$, $2\leq s\leq d-1$, for $i,j=1,\ldots,k$.
   \begin{enumerate}

           \item Let $B=A'_TA_T-I$, derive
$$\lambda_k(B) \geq
  \left\{
   \begin{array}{ll}
   -sk/2d&if~k~is~even  \\[6pt]
   -s\sqrt{k^2-1}/2d & if~k~is~odd\\
   \end{array}
  \right.,
$$ and then  $$\lambda_k(A'_TA_T)=1+\lambda_k(B)\geq 1-\frac{sk}{2d}.$$
           \item Let $C=A'_TA_T-(1-\frac{s}{d})I$, derive $\lambda_1(C)\leq ks/d$, and then $$\lambda_1(A'_TA_T)\leq \frac{(k-1)s+d}{d}.$$
         \end{enumerate}

   \item if $M/2<d\leq M-1$, $[A'_TA_T]_{ii}=1$ and $[A'_TA_T]_{ij, i\neq j}$ $\in  \{(2d-M)/d,\ldots,s/d\}$,  $2d-M\leq s\leq d-1$, for $i,j=1,\ldots,k$.
    \begin{enumerate}
    \item  Let $B=A'_TA_T-(1-\frac{2d-M}{d})I$, derive

$$
\lambda_k(B) \geq
  \left\{
   \begin{array}{ll}
   \frac{ k(2d-M-s)}{2d}&\footnotesize{if~k~is~even}  \\[6pt]
   \frac{k(2d-M)-\sqrt{(2d-M)^2-(k^2-1)s^2}}{2d}&\footnotesize{if~k~is~odd}\\
   \end{array}
  \right.,
$$\\ \vspace{10pt}
further deduce $\lambda_k(B)\geq -\frac{k(2d-M-s)}{2d}$, and then we have that
$$\lambda_k(A'_TA_T) \geq \frac{k(2d-M-s)+2(M-d)}{2d}$$
\item Let $C=A'_TA_T-(1-\frac{s}{d})I$, derive $\lambda_1(C)\leq ks/d$, and then it follows that\\ $$\lambda_1(A'_TA_T)\leq \frac{(k-1)s+d}{d}.$$
    \end{enumerate}

 \item Finally, it follows from $\delta_k=\frac{\lambda_1-\lambda_k}{\lambda_1+\lambda_k}$ that\\
 \\
 $$
 \large{\delta_k} =
\left\{
   \begin{array}{ll}
   %\begin{subequations}
\frac{\textstyle (3k-2)s}{\textstyle (k-2)s+4d}&\footnotesize{if~ 3\leq d\leq \frac{M}{2}}~ and ~\footnotesize{2\leq s\leq d-1} \\[12pt]
\frac{\textstyle (3k-2)s+(k-2)(M-2d)}{\textstyle (k-2)s-(M-2d)k+2M}&\footnotesize{if ~\frac{M}{2}< d\leq M-2} ~and~ \footnotesize{2d-M\leq s\leq d-1} \\
  %\end{subequations}
   \end{array}
   \right.$$

 \end{enumerate}
 \end{IEEEproof}
%\section*{Acknowledgment}

%The authors would like to thank...

% Can use something like this to put references on a page
% by themselves when using endfloat and the captionsoff option.
\ifCLASSOPTIONcaptionsoff
  \newpage
\fi

% trigger a \newpage just before the given reference
% number - used to balance the columns on the last page
% adjust value as needed - may need to be readjusted if
% the document is modified later
%\IEEEtriggeratref{8}
% The "triggered" command can be changed if desired:
%\IEEEtriggercmd{\enlargethispage{-5in}}

% references section

% can use a bibliography generated by BibTeX as a .bbl file
% BibTeX documentation can be easily obtained at:
% http://www.ctan.org/tex-archive/biblio/bibtex/contrib/doc/
% The IEEEtran BibTeX style support page is at:
% http://www.michaelshell.org/tex/ieeetran/bibtex/
%\bibliographystyle{IEEEtran}
% argument is your BibTeX string definitions and bibliography database(s)
%\bibliography{IEEEabrv,../bib/paper}
%
% <OR> manually copy in the resultant .bbl file
% set second argument of \begin to the number of references
% (used to reserve space for the reference number labels box)

\bibliographystyle{IEEEtran}
\bibliography{egbib}
% biography section
%
% If you have an EPS/PDF photo (graphicx package needed) extra braces are
% needed around the contents of the optional argument to biography to prevent
% the LaTeX parser from getting confused when it sees the complicated
% \includegraphics command within an optional argument. (You could create
% your own custom macro containing the \includegraphics command to make things
% simpler here.)
%\begin{IEEEbiography}[{\includegraphics[width=1in,height=1.25in,clip,keepaspectratio]{mshell}}]{Michael Shell}
% or if you just want to reserve a space for a photo:

%\begin{IEEEbiography}{Michael Shell}
%Biography text here.
%\end{IEEEbiography}

% if you will not have a photo at all:
%\begin{IEEEbiographynophoto}{John Doe}
%Biography text here.
%\end{IEEEbiographynophoto}

% insert where needed to balance the two columns on the last page with
% biographies
%\newpage

%\begin{IEEEbiographynophoto}{Jane Doe}
%Biography text here.
%\end{IEEEbiographynophoto}

% You can push biographies down or up by placing
% a \vfill before or after them. The appropriate
% use of \vfill depends on what kind of text is
% on the last page and whether or not the columns
% are being equalized.

%\vfill

% Can be used to pull up biographies so that the bottom of the last one
% is flush with the other column.
%\enlargethispage{-5in}

% that's all folks

\end{document}